\documentclass[aps,prx,twocolumn, superscriptaddress,amsmath,amssymb,reprint,footinbib,longbibliography]{revtex4-1}
\usepackage{amsfonts}
\usepackage{amsmath}
\usepackage{graphicx}
\usepackage{amssymb}
\usepackage{physics}
\usepackage{bm}
\usepackage{mathptmx}
\usepackage{xcolor}
\usepackage{listings}
\usepackage[breaklinks=true,colorlinks=true,linkcolor=magenta,urlcolor=blue,citecolor=blue]{hyperref}

\definecolor{mygreen}{RGB}{28,172,0} % color values Red, Green, Blue
\definecolor{mylilas}{RGB}{170,55,241}

\begin{document}
    \title{Charge noise and overdrive errors in reflectometry-based charge, spin and 
    Majorana qubit readout}
    \author{Vahid Derakhshan Maman}
    \email{vahid@phy.bme.hu}
    \affiliation{Department of Theoretical Physics and
        MTA-BME Exotic Quantum Phases "Momentum" Research Group, Budapest University of Technology and Economics, 
        H-1111 Budapest, Hungary}
    
    \author{M.~F.~Gonzalez-Zalba}
    \email{mg507@cam.ac.uk}
    \affiliation{Hitachi Cambridge Laboratory, J. J. Thomson Ave., Cambridge, CB3 0HE, United Kingdom}

    \author{Andr\'as P\'alyi}
    \email{palyi@mail.bme.hu}
    \affiliation{Department of Theoretical Physics and
        MTA-BME Exotic Quantum Phases "Momentum" Research Group, Budapest University of Technology and Economics, 
        H-1111 Budapest, Hungary}
    
\date{\today}  
    
\begin{abstract}
    
    Solid-state qubits incorporating quantum dots can be read out by gate reflectometry.
    Here, we theoretically describe physical mechanisms that render such 
    reflectometry-based readout
    schemes imperfect.
    We discuss charge qubits, singlet-triplet spin
    qubits, and Majorana qubits.
    In our model, we account for readout errors 
    due to slow charge noise, and due to overdriving, 
    when a too strong probe is causing errors.
    A key result is that for charge and spin qubits, 
    the readout fidelity saturates at large 
    probe strengths, whereas for Majorana qubits, 
    there is an optimal probe strength which provides a 
    maximized readout fidelity.
    We also point out the existence of severe readout errors
    appearing in a resonance-like fashion as the 
    pulse strength is increased, and show that 
    these errors are related to probe-induced multi-photon 
    transitions.
    Besides providing practical guidelines toward optimized 
    readout,
    our study might also inspire 
    ways to use gate reflectometry for device characterization. 
    
%    In these notes, we theoretically study error mechanisms
%        in charge-qubit readout based on reflectometry. 
%        In particular, we describe overdrive effects, 
%        i.e., a reduction of readout fidelity as the strength of the
%        probe pulse is increased. 
%        We point out the existence of severe readout errors
%        appearing in a resonance-like fashion as the 
%        pulse strength is increased, and show that 
%        these errors are related to efficient multi-photon 
%        transitions.
%        To quantify readout fidelity, we also account for 
%        amplifier noise.
\end{abstract}
\maketitle

%\tableofcontents

    \section{Introduction}
    
%    Topological quantum computation is an approach to storing and manipulating quantum information in which the unitary quantum gates result from certain topological quantum objects \cite{Kitaev2003}.  Quantum version of the classical bit, the quantum bit is a key ingredient for quantum computing. Previously, a few various physical realizations of qubits have been suggested and proposed experimentally, including superconducting circuits, \cite{Nori-2011, Kockum-2017} trapped ions,\cite{Monroe-2013} donors in silicon, \cite{Dehollain-2012} semiconductor quantum dots, \cite{Veldhorst-2015, Yao-2012}, etc. Among them, semiconductor quantum dots, due to compatibility with the metal-oxide-semiconductor technology  are most promising. During the last couple of decades, qubits, by pursuing degrees of electrons or holes in quantum dots have been developed, including charge states of electrons, \cite{Hayashi-2003} spin states of an electron \cite{Loss-1998} or hole, \cite{Petta-2009} singlet-triplet states of two electrons, and other hybrid states, \cite{DiVincenzo-2000, Simmons-2014}.
    
%One significant challenge facing all of these architectures is the loss of coherence due to interactions with environmental degrees of freedom.
    
Readout of quantum bits is an essential ingredient in 
practical quantum computing.
For quantum-dot charge qubits and spin qubits,
a natural way of readout is based on charge measurements \cite{HansonRevModPhys2007}.
An alternative is to measure charge susceptibility, 
that is, the ability of a charge to be displaced when 
subject to a changing electrostatic environment, typically 
the gate voltage of a nearby gate electrode.
Recent works are pointing to the possibility of 
applying those measurements
to hybrid superconductor-semiconductor devices, 
potentially serving as topological quantum bits 
based on Majorana zero 
modes 
\cite{KarzigPRB2017,PluggeNewJPhys2017,vanVeenPRB2019, deJongPRApplied2019, RazmadzePRApplied2019,
SabonisAPL2019}.

\begin{figure}
    \begin{center}
        \includegraphics[width=1\columnwidth]{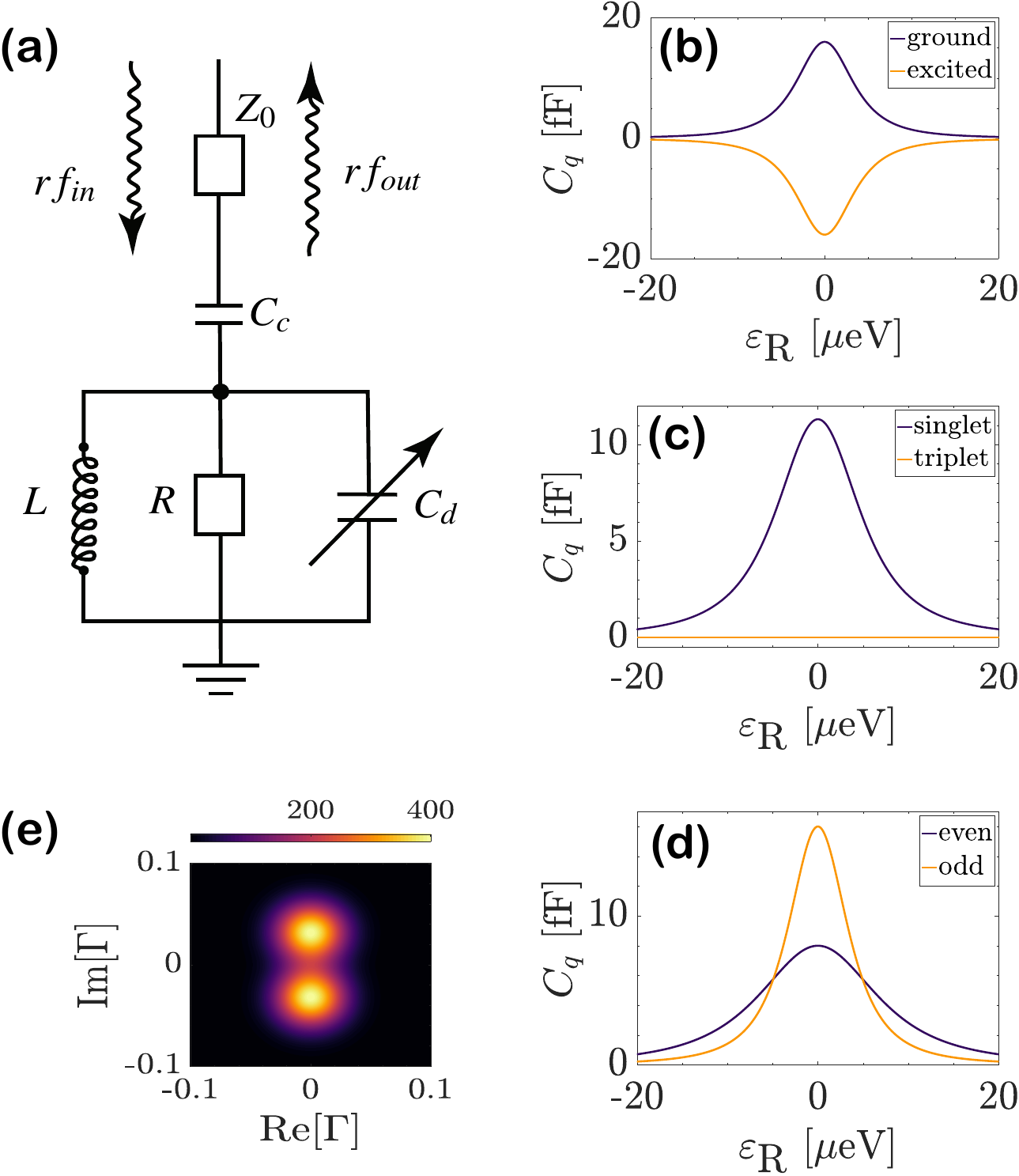}
    \end{center}
    \caption{Dispersive readout of charge, spin and Majorana qubits.
    (a) Typical circuit. 
    The parametric capacitance $C_q$
    depending on the qubit state affects $C_d = C_0 + C_q$,
    hence changes the ac response of the resonant circuit.
    (b) $C_q$ of the ground (purple) and
    excited (orange) state of a charge qubit [Eq.~\eqref{eq:chargequbitcapacitance}], 
    for $\Delta = 5 \, \mu$eV.
    (c) $C_q$ of the singlet (purple) 
    and triplet (orange)
    state of a singlet-triplet spin qubit
    [Eq.~\eqref{eq:singletcapacitance}],
    for $\Delta = 5 \, \mu$eV. 
    (d) $C_q$ of the even (purple) and odd (orange) states
    of a Majorana qubit [Eq.~\eqref{eq:majoranacapacitance}],
    with $\Delta_\text{even} = 10 \, \mu$eV
    and $\Delta_\text{odd} = 5 \, \mu$eV.
   (e) Reflectance probability distributions [Eq.~\eqref{eq:gammagpdf}] due to
    amplifier noise for the ground (upper blob)
    and excited (lower blob) 
    charge-qubit states, according to 
    Eq.~\eqref{eq:gammagpdf}.
    Parameters: tunnel splitting: $\Delta = 5\, \mu$eV,
    dimensionless amplifier noise strength $\sigma_\Gamma = 0.02$,
    linear circuit response coefficient: $\alpha = 0.002 \, \text{fF}^{-1}$.
    Overdrive effects are excluded.
    \label{fig:preliminaries}}
    \end{figure}

Here, we provide a theoretical analysis
of qubit readout errors specific to the 
readout method known as reflectometry-based 
dispersive readout.
In this method, the device hosting the quantum bit
is embedded in a classical resonant circuit,
and the impedance of the overall system is 
probed in a reflection as shown in 
Fig.~\ref{fig:preliminaries}a.
The qubit influences the resonator impedance 
by changing its total capacitance $C_d = C_0 + C_q$, 
where $C_0$ is the resonator capacitance and $C_q$ is the qubit-state-dependent 
parametric capacitance of a quantum dot that
is either part of the qubit (charge qubit, spin qubit)
or used as an auxiliary element enabling readout (Majorana qubit). 

We describe generic error mechanisms that influence
the readout of charge, spin, and Majorana qubits:
amplifier noise, charge noise, and overdrive effects.
Besides establishing simple models of these mechanisms, 
and analyzing their consequences for qubit readout,
we also offer practical guidelines to optimize readout
precision. 
Our results might also inspire new ways to
utilize reflectometry for device characterization. 

The rest of the paper is structured as follows.
In section \ref{sec:preliminaries}, we introduce
charge, spin and Majorana qubits, 
and review how amplifier noise reduces the readout
fidelity in a gate-based dispersive readout experiment.
In section \ref{sec:chargenoise}, we describe the effect
of charge noise. 
In section \ref{sec:adiabatic}, we describe how overdriving
affects the readout fidelity in a simplified,
adiabatic model. 
In section \ref{sec:transitions}, we go beyond the 
adiabatic description and discuss the resonant reduction 
of readout fidelity originating from probe-induced
multi-photon transitions in the device.
Section \ref{sec:discussion} provides a 
discussion, including open problems, 
and the paper is concluded in section \ref{sec:conclusions}.
Details not required to follow the main text 
are delegated to the appendices.

\section{Preliminaries}
\label{sec:preliminaries}

\subsection{Charge qubit}
\label{sec:chargequbit}

Here, we describe a typical gate-based dispersive readout setup, 
and exemplify how it works with the example of a double-dot
charge qubit \cite{XiaoMiScience2017,ScarlinoPRL2019,
MizutaPRB2017,AhmedPhysRevApplied2018, CollessPRL2013,Gonzalez-Zalba2016,StockklauserPRX2017,ChatterjeePRB2018}.
The readout circuit is shown in Fig.~\ref{fig:preliminaries}a.
A radiofrequency signal of amplitude $V_\text{in}$ 
and frequency $\omega/(2\pi)$
is sent through a transmission line ($Z_0$)
to the resonator, where it is reflected, 
due to impedance mismatch, producing an output signal
$V_\text{out}=\Gamma V_\text{in}$, 
where $\Gamma$ is the reflection coefficient
(which we will call `reflectance' for short).
$V_\text{out}$ is measured, typically after cryogenic amplification, 
with an IQ demodulation system.
The reflectance can be expressed as 
$\Gamma(\omega)=(Z(\omega)-Z_0)/(Z(\omega)+Z_0)$, where 
$Z(\omega)$ is the 
frequency-dependent equivalent impedance of the resonator,
and its absolute value is limited: 
$|\Gamma| \leq 1$.

The reflectance carries information about the qubit 
state through the state-dependent parametric
capacitance component $C_q$ of the device.
In an ideal scenario, each of the two qubit states yields
well-defined and distinct 
reflectance values, and hence this measurement
allows for perfect qubit readout.
However, non-idealities, such as amplifier noise
\cite{StehlikPRApp2015,SchaalPRL2020} or
on-device charge noise \cite{PetitPRL2019} 
render reflectance a random 
variable, and hence make the inference of the qubit state
from the measured reflectance imperfect.

A simple model of the charge qubit
is as follows. 
The qubit is formed by a single electron 
confined in a double quantum dot, described by
the $2\times 2$ Hamiltonian matrix
\begin{equation}
\label{eq:chargequbithamiltonian}
    H = 
    %\frac{\Delta}{2} \sigma_x 
    %+ \frac{\varepsilon_\text{R}}{2} \left(1-\sigma_z\right),
    \left(\begin{array}{cc}
    0 & \Delta/2 \\
    \Delta/2 & \varepsilon_\text{R}
    \end{array}\right),
\end{equation}
where $\Delta$ is the tunneling amplitude between the 
two dots, and $\varepsilon_\text{R}$ is the detuning of the
on-site energy of the right dot from 
that of the left dot. 

To mathematically describe the readout process we make
assumptions (i)-(vi) below.
None of these are critical for the
conclusions of our paper, we apply them only to keep
the discussion focused on the most important
physical effects.
For more detailed discussion, see
section \ref{sec:discussion} and 
appendix \ref{app:circuit}.

(i) The resonator is directly connected to the gate electrode of the right dot.

(ii) Among the mutual capacitances of the right dot 
and its surrounding metallic elements (including 
the left dot), the capacitance 
with its plunger gate dominates. 
This implies that the so-called `lever arm' parameter 
$\kappa$, describing the coupling of the right 
dot with its gate, is well approximated
as $\kappa = 1$.

(iii) The circuit is probed at its eigenfrequency 
corresponding to zero parametric capacitance.

(iv) On the one hand, the integration time is much longer then
a single period of the probe pulse, and the transient
time scales of the resonator.
On the other hand, the integration time is much shorter than the
characteristic relaxation times of the system.

(v) 
For simplicity, we assume that the resonator
is perfectly impedance-matched to the 
transmission line at its resonance frequency
(at zero parametric capacitance).

(vi)
Furthermore, we assume that
the qubit-state-dependent parametric capacitance 
of the device is small, in the sense that
the change in reflectance 
due to a change in the parametric
capacitance is well approximated by a linear dependence.
Together with condition (v), this implies
%\begin{equation}
%\label{eq:linearresponse}
%\Gamma =i\frac{2\beta}{(1+\beta)^2}Q_0\frac{C_\text{q}}{C_\%text{0}+C_\text{c}}= i \alpha C_\text{q},
%\end{equation}
\begin{equation}
\label{eq:linearresponse}
\Gamma =
     i \alpha C_q,
\end{equation}
where $\alpha > 0$ is a coefficient with 
dimension capacitance$^{-1}$, whose exact value is
determined by the electrical parameters of the resonator,
see Appendix \ref{app:circuit} for an example.
%such as the internal quality factor $Q_\text{0}$ and the 
%matching coefficient 
%$\beta=Z_0/Z_\text{eq}$~\cite{IbbersonAPL2019}. 
%Condition (v) implies, we consider the small signal limit, 
%where the changes in quantum capacitance are small, which 
%in the case of a critically coupled resonator corresponds 
%to $C_\text{q}\ll 2(C_\text{0}+C\text{c})/Q_\text{l}$ where
%$Q_\text{l}$ is the loaded quality factor of the resonator.
Note that even without perfect impedance matching, 
a linear relation between $\Gamma$ and $C_q$ could
be established, with a slightly modified
form $\Gamma = \alpha_0 + \alpha_1 C_q$ with
complex-valued $\alpha_0$ and $\alpha_1$;
nevertheless, we use the 
simpler Eq.~\eqref{eq:linearresponse}
throughout this paper.

For the charge qubit, we can associate a
parametric capacitance $C_q$ 
to the ground state $g$
\cite{Petersson2010aNanolett,MizutaPRB2017},
and one to the excited state $e$:
\begin{equation}
\label{eq:chargequbitcapacitance}
    C_{q,s} = \sigma_s \, e^2
\frac{\Delta^2/2}{(\varepsilon_\text{R}^2 + \Delta^2)^{3/2}},
\end{equation}
where $s\in\{g,e\}$ and
$\sigma_g = +1$ and $\sigma_e = -1$.
These parametric capacitances are plotted in 
Fig.~\ref{fig:preliminaries}b, as functions 
of  detuning $\varepsilon_\text{R}$.
Note that the peak parametric capacitance 
for a typical value of $\Delta = 5\, \mu$eV 
at the tipping point $\varepsilon_\text{R} = 0$ is
$C_{q,g} \approx 16$ fF.

Assuming an ideal, noiseless scenario, where the 
readout point is the tipping point $\varepsilon_\text{R} = 0$,
and assuming that the quantum state to be observed
is either $g$ or $e$, 
the experimenter will measure one of the two
possible reflectance values,
$\Gamma_g = i \alpha C_{q,g}$ or 
$\Gamma_e = i \alpha C_{q,e}$. 
In the presence of amplifier noise, the 
detected reflectance becomes a random variable, 
with a Gaussian noise added to both quadratures.
Hence
the probability density function (pdf) of the 
reflectance for the state $g$ reads
\begin{equation}
\label{eq:gammagpdf}
    P_g(\text{Re}(\Gamma),\text{Im}(\Gamma)) = 
    G(\text{Re}(\Gamma),\sigma_\Gamma)
%    G(\text{Im}(\Gamma)-\Gamma_g,\sigma_\Gamma),
    G(\text{Im}(\Gamma)-i \alpha C_{q,g},\sigma_\Gamma),
\end{equation}
where $G(x,\sigma)$ is the pdf of a Gaussian random variable $x$
with zero mean and standard deviation $\sigma$, and
$\sigma_\Gamma$ is the dimensionless amplifier noise strength
(see below).
For the state $e$, the reflectance pdf $P_e(\Gamma)$
has a form analogous to Eq.~\eqref{eq:gammagpdf}.
The two-dimensional Gaussian pdfs
corresponding to the states $g$ and $e$
are shown in Fig.~\ref{fig:preliminaries}e.

We define the dimensionless amplifier 
noise strength 
$\sigma_\Gamma$ as the square root of the ratio of the
amplifier noise power $P_\text{n}$ 
and the input power at the resonator 
$P_\text{in}=V_\text{in}^2/Z_0$
\cite{Muller1974, SchaalPRL2020}. 
The latter can be expressed in terms of the
ac voltage amplitude
at the gate of the right dot, 
$V_\text{dev}$, giving:
\begin{equation}
\label{eq:sigmagamma}
\sigma_\Gamma = \sqrt{\frac{P_\text{n}}{P_\text{in}}} = \frac{\sqrt{k_\text{B} T_\text{n} R}}{V_\text{dev} \sqrt{t_\text{int}}},
\end{equation}
where $k_\text{B}$ is Boltzmann's constant,
$T_\text{n}$ is the effective noise temperature of the 
amplifier, 
$R$ is the resistance in the
resonator (see Fig.~\ref{fig:preliminaries}a), 
and $t_\text{int}$ is the integration time. 

Since the reflectance is bounded, $|\Gamma| \leq 1$, 
a reasonable experimental setup is expected to 
have $\sigma_\Gamma \lesssim 1$.
% For example, for $T_\text{n} = 4$ K, 
% $R = 50\, \Omega$,
% $V_\text{dev} = 1\, \mu$V, and
% $t_\text{int} = 1 \mu$s, 
% we have $\sigma_\Gamma \approx 0.0525$.
For example, for $T_\text{n} = 0.1$ K corresponding
to a Josephson parametric amplifier, 
$R = 577\, \text{k}\Omega$ (see example circuit in 
appendix \ref{app:circuit}),
$V_\text{dev} = 10 \, \mu\text{eV}$
and $t_\text{int} = 1 \, \mu\text{s}$,
we have $\sigma_\Gamma \approx 0.126$.
%Equation \eqref{eq:sigmagamma} and its parameters
%are further discussed in Appendix [AmplifierNoise].

The randomness of the measured reflectance implies
that the experimenter can make an erroneous inference
of the qubit state based on the measured reflectance. 
We will characterize this readout error 
by two different quantities:
the \emph{signal-to-noise ratio} (SNR), which
we apply only to cases where the two 
reflectance distributions are circularly
symmetric Gaussians and
have the same standard deviations, 
and the \emph{readout fidelity}, which we apply
generically.

As long as the measurement noise is dominated by 
amplifier noise, and quantum-mechanical 
squeezing effects\cite{ClerkRevModPhys2010} can be disregarded,
then it is guaranteed that the two reflectance pdfs 
$P_g(\Gamma)$ and $P_e(\Gamma)$ are
circularly symmetric two-dimensional Gaussians with identical
standard deviation, as depicted in Fig.~\ref{fig:preliminaries}e. 
In this case, the probability of correct inference
of the qubit state increases monotonically with 
increasing SNR, where SNR is defined as
\begin{equation}
\label{eq:snrdef}
    \text{SNR} = \frac{|\Gamma_g - \Gamma_e|}{\sigma_\Gamma}.
\end{equation}
For example, we have $\text{SNR} \approx 3.2$, if
we take the dimensionless amplifier noise strength
$\sigma_\Gamma = 0.02$ 
and parametric capacitance $C_{q,g} = 16\, \text{fF}$, 
and set the circuit's linear response 
coefficient to $\alpha = 0.002 \text{fF}^{-1}$,
as shown in Fig.~\ref{fig:preliminaries}e.

We will see that charge noise can distort the reflectance
distributions of
$\Gamma_g$ and $\Gamma_e$, and then
the simple SNR definition in Eq.~\eqref{eq:snrdef} 
is not applicable.
In those cases, we will use the readout fidelity,
whose definition is as follows. 
The starting point is the
maximum likelihood inference rule.
This inference rule assumes that the experimenter
knows the reflectance pdfs for both qubit states from
a model, 
and when she measures a certain reflectance value, 
then she will attribute it to the qubit state that has 
the higher
probability of producing this reflectance value. 
Given this inference rule, the readout 
fidelity $F$ is defined as the probability of correct state 
inference, 
assuming that the experimenter has zero a priori knowledge
of the state.
(See appendix \ref{app:maximumlikelihood} for a more
detailed
discussion.)

In the presence of Gaussian amplifier noise as described
above, this principle translates to the following integral:
\begin{equation}
\label{eq:fidelity}
    F = \int_{-\infty}^{\infty} d \Gamma'
    \int_0^\infty d\Gamma'' P_g(\Gamma',\Gamma'')
    = \frac{1}{2}
    \left[
    1+ \text{erf}\left(\frac{\text{SNR}}{2\sqrt{2}}\right)
    \right].
\end{equation}
Here, erf denotes the error function, which is a monotonically
increasing function. 
In Eq.~\eqref{eq:fidelity},
the integral of the reflectance imaginary part starts
from zero, since, e.g., if the qubit is in state $g$, 
and we measure a reflectance with a negative 
imaginary part, then the inferred
state is $e$ and that is counted as a readout error.
An example: 
with the parameter values below
Eq.~\eqref{eq:snrdef}, 
yielding $\text{SNR} \approx 3.2$,
the readout fidelity is $F \approx 0.945$.

\subsection{Spin qubit}

Now we illustrate
reflectometry-based qubit readout, and the harmful effect
of ampifier noise, on the example of a singlet-triplet spin 
qubit.
For concreteness, we take the two-electron
qubit based on the singlet ground state $S_g$
and the unpolarized triplet $T_0$ \cite{PettaScience2005,HansonRevModPhys2007}.
In this encoding, the parametric
capacitance of the state $T_0$ is zero,
$C_{q,T} = 0$.
Furthermore, the effective Hamiltonian of the singlet bonding
and antibonding states in the vicinity of the (1,1) - (0,2) 
tipping point reads
\begin{equation}
\label{eq:singlethamiltonian}
    H = \left(\begin{array}{cc}
        U+ \varepsilon_\text{R} & \Delta/\sqrt{2} \\
        \Delta/\sqrt{2} & U + 2\varepsilon_\text{R}
    \end{array}\right),
\end{equation}
where 
$\Delta$ is the single-electron hybridization gap, 
$U$ denotes the on-site Coulomb repulsion energy,
$\varepsilon_\text{R}$ denotes the on-site
energy of the right dot,
the on-site energy of the left dot is set to
$\varepsilon_\text{L} = U$
for convenience, and 
the basis state ordering
in Eq.~\eqref{eq:singlethamiltonian} is (1,1), (0,2).
Note that the (1,1)-(0,2) hybridization gap is
$\Delta' = \sqrt{2} \Delta$.

Then, the parametric capacitance of the singlet
ground state $S_g$ is (cf. the relation between 
Eqs.~\eqref{eq:chargequbithamiltonian} 
and \eqref{eq:chargequbitcapacitance})
\begin{equation}
\label{eq:singletcapacitance}
C_{q,S} =  e^2
\frac{\Delta^2}{(\varepsilon_\text{R}^2 + 2 \Delta^2)^{3/2}}.   
\end{equation}
The parametric capacitances $C_{q,S}$ and $C_{q,T}$ 
as functions of the 
detuning parameter $\varepsilon_\text{R}$
are shown in 
Fig.~\ref{fig:preliminaries}c.

The parametric capacitance of each qubit basis state
determines the corresponding reflectance
$\Gamma_T$ and $\Gamma_S$ via the
linear circuit response approximation in Eq.~\eqref{eq:linearresponse}.
In the presence of amplifier noise, 
the SNR is defined in analogy with the
charge qubit (Eq.~\eqref{eq:snrdef}) as 
$\text{SNR} = |\Gamma_S - \Gamma_T|/\sigma_\Gamma$.
Then, the readout fidelity formula Eq.~\eqref{eq:fidelity}
remains unchanged.

\subsection{Majorana qubit}

Finally, we discuss the case of a Majorana 
qubit
\cite{Kitaev2001unpaired,AliceaNatPhys2011,KarzigPRB2017,PluggeNewJPhys2017,FlensbergPRL2011,GharaviPRB2016,KnappPRB2018,MunkArxiv2020,SteinerArxiv2020}.
For concreteness, we follow the description in Ref.~\cite{FlensbergPRL2011},
but the results generalize naturally to more recent
Majorana qubit proposals, e.g., the
tetron
\cite{KarzigPRB2017,PluggeNewJPhys2017}, 
where charging energy protects the qubits from 
quasiparticle poisoning. 

In the setup of Ref.~\cite{FlensbergPRL2011}, 
two Majorana zero modes $\gamma_1$ and $\gamma_2$
are coupled to an auxiliary quantum dot. 
This setup can be used to read out the joint parity of the
two Majorana zero modes. 
The joint Majorana parity can also be expressed as
the occupation
$d^\dag d$
of the nonlocal fermion $d = (\gamma_1 + i \gamma_2)/2$,
so that the low-energy dynamics of the system 
can be studied using the occupation number basis 
of the readout dot and the non-local fermion, 
$\ket{n_d,n_f}$, where $n_d, n_f \in \{0,1\}$.

The complex-valued tunnel coupling between the dot and
$\gamma_1$ ($\gamma_2$) is denoted as $v_1$ ($v_2$),
following Ref.~\cite{FlensbergPRL2011}.
These couplings can be tuned
by adjusting barrier 
gate voltages and an Aharanov-Bohm flux
piercing the device. 
The effective low-energy Hamiltonian for the even 
and odd sectors read
\begin{equation}
\label{eq:majoranahamiltonian}
    H_\tau= \frac{1}{2}\left(\begin{array}{cc}
    0 & \Delta_\tau \\
    \Delta_\tau & \varepsilon_\text{R}
    \end{array}\right), \, \,  \tau \in \{\text{even},\text{odd}\},
\end{equation}
where $\Delta_\text{even} = 2|v_1 - i v_2|$
and $\Delta_\text{odd} = 2|v_1 + i v_2|$,
$\varepsilon_\text{R}$ is the on-site energy of the
readout dot,
and the basis states are 
$(e^{i \arg(v_1-iv_2)}\ket{0_d,0_f},\ket{1_d,1_f})$ for the even 
case,
and 
$(e^{i \arg(v_1+i v_2)}\ket{1_d,0_f},\ket{0_d,1_f})$ for the odd 
case.
The only difference between our 
Eq.~\eqref{eq:majoranahamiltonian}, and
Eq.~(7) in Ref \cite{FlensbergPRL2011},
is that we 
have done the above basis transformation 
to make the tunnel matrix elements positive ($\Delta_\tau > 0$).

As long as the tunnel couplings $v_1$ and $v_2$
are not fine-tuned, the matrix elements $\Delta_\text{even}$
and $\Delta_\text{odd}$ are different, and the
even and odd cases can be distinguished by measuring any quantity
depending on $\Delta_\tau$, e.g., the
parametric capacitance of the readout dot. 
This is the readout scheme we describe here.
In practice, for example, one could start with 
the qubit decoupled
from the readout dot ($v_1 = v_2 = 0$) and with positive
dot energy $\varepsilon_\text{R}$, 
either in state $\ket{0_d,0_f}$ or $\ket{1_d,0_f}$. 
Then, one can gradually (e.g., adiabatically)
turn on the qubit-dot coupling and lower the dot energy
such that the Hamiltonian in Eq.~\eqref{eq:majoranahamiltonian}
is reached, and the state of the qubit-dot system reaches
the ground state of the corresponding Hamiltonian. 
Then, reflectometry measures the ground-state
parametric capacitance
\begin{equation}
\label{eq:majoranacapacitance}
C_{q,\tau} = e^2
\frac{\Delta_\tau^2/2}{(\varepsilon_\text{R}^2(t) + \Delta_\tau^2)^{3/2}}.
\end{equation}
These capacitances are shown in Fig.~\ref{fig:preliminaries}d,
as functions of the readout dot on-site energy $\varepsilon_\text{R}$,
with $\Delta_\text{even} = 10\,\mu$eV
and $\Delta_\text{odd} = 5\, \mu$eV.

The parametric capacitances determine the reflectances 
$\Gamma_\text{even}$ and $\Gamma_\text{odd}$ via the
linear circuit response approximation in Eq.~\eqref{eq:linearresponse},
implying 
$\text{SNR} = |\Gamma_\text{even} - \Gamma_\text{odd}| /\sigma_\Gamma$,
and with this SNR, 
the fidelity is still expressed as in Eq.~\eqref{eq:fidelity}.

\section{Charge noise}
\label{sec:chargenoise}

In real qubit devices, the electromagnetic environment is not 
fully controlled, and hence
the qubit is subject to random electromagnetic fluctuations.
In many quantum-dot experiments, experimental features \cite{FreemanAPL2016,YonedaNatNano2018,BoterArxiv2019}
are well explained by the model that the quantum-dot on-site energies
slowly fluctuate, e.g., 
following a $1/f$-type power spectrum 
\cite{FreemanAPL2016,YonedaNatNano2018}.

Remarkably, often the noise-induced features are captured
by a \emph{quasistatic} noise model. 
There, it is assumed that
the on-site energies of the quantum dots change between subsequent
measurement cycles, but
are static within a single measurement cycle
with characteristic duration $t_\text{int}$. 
This is a convenient, popular, and fruitful model
for typical quantum-information experiments where
the experimenter has to carry out many measurement cycles
to obtain a statistically accurate description of the qubit's
dynamics. 
The success of this quasistatic approximation
can be attributed to the fact that the power spectrum 
of $1/f$-type noise is indeed dominated by low frequencies.

Following earlier work \cite{XinWang2012NatComm,TosiNatComm2017, WatsonNature2018, BoterArxiv2019,DovzhenkoPRB2011}, we will adopt the quasistatic
description of charge noise in this section, 
and apply it to 
describe errors in reflectometry-based
readout of the charge, spin and Majorana qubits 
introduced above.
On the one hand, this will provide guidelines how
to mitigate the noise effects that are harmful to 
qubit readout. 
On the other hand, this analysis can also be interpreted
as a proposal to use the qubits as tools for 
charge-noise characterization. 

We consider the charge qubit, add charge noise, but first, 
we disregard amplifier noise. 
Charge noise is then modelled by random 
components of the on-site energies of the two dots, 
which are assumed to be quasistatic, Gaussian, and independent
of each other. 
Each random on-site-energy component has a standard deviation
$\sigma_c$.
This is equivalent to disregarding the random component
of $\varepsilon_\text{L}$ but taking a random component
$\delta \varepsilon$ on $\varepsilon_\text{R}$
with standard deviation $\sigma_\varepsilon = \sqrt{2} \sigma_c$.
According to charge-qubit
experiments \cite{ScarlinoPRL2019, PettaPRL2004,PeterssonPRL2010,DovzhenkoPRB2011,ShiPRB2013,ThorgrimssonNPJQuantumInfo2017}, 
the typical value
of this charge noise strength in state-of-the-art
devices ranges between
$\sigma_\varepsilon \sim 0.2 \, \mu$eV and
$\sigma_\varepsilon \sim 10 \, \mu$eV.
Due to the random character of the on-site energies, 
the parametric capacitance will also 
be random, albeit not Gaussian, since
the relation \eqref{eq:chargequbitcapacitance} 
between the detuning $\varepsilon_\text{R}$ and
the capacitance is not linear. 
In turn, the randomness of the parametric capacitance
will carry over to the reflectance in a simple, linear
fashion, at least provided that our
linear-response assumption in Eq.~\eqref{eq:linearresponse}
holds. 

    \begin{figure*}
    \begin{center}
    \includegraphics[width=1.7\columnwidth]{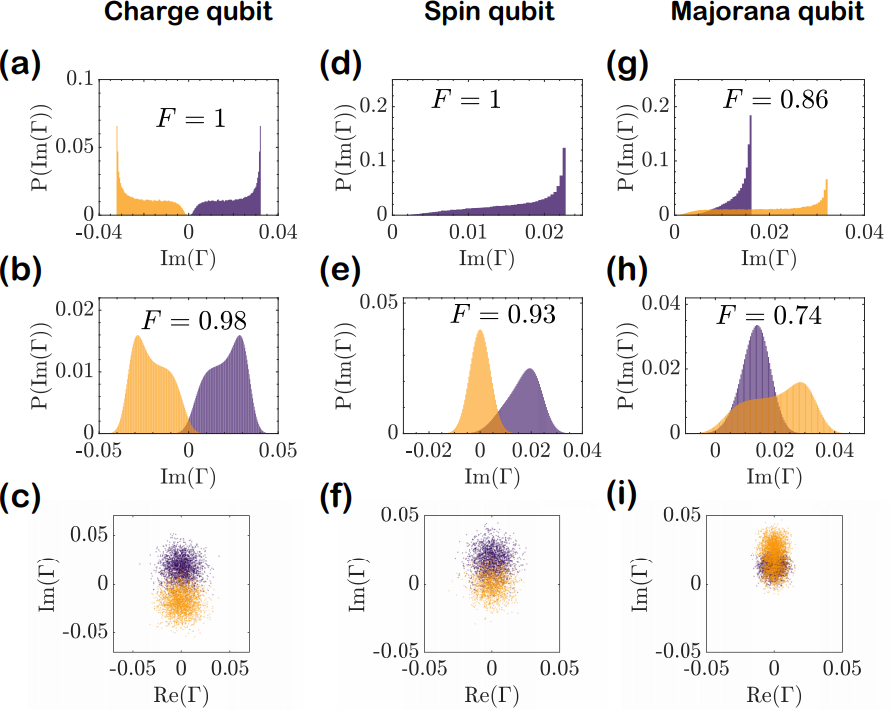}
    \end{center}
    \caption{Effect of charge noise and amplifier
    noise on the readout of charge, spin and 
    Majorana qubits.
    (a) Histogram of reflectance associated to the ground (purple) and excited (orange) states 
    for the case of charge qubit when only the charge noise is present.
    (b) Reflectance histogram for a charge qubit, 
    corresponding to the ground (blue) and excited (orange) states,
    in the presence of both charge noise and amplifier noise. 
    (c) Scatter plots showing the random reflectance values.
    (d) Spin-qubit reflectance histogram of the singlet state, 
    with charge noise only.
    (The triplet case is trivial, $\text{Im}(\Gamma) = 0$
    for all realizations, hence it is not shown.)
    (e) Spin-qubit reflectance histogram for the singlet (purple) and
    triplet (orange) states, with charge noise and amplifier noise.
    (f) Scatter plots showing the random reflectance values for the two spin-qubit basis
    states. 
    (g) Majorana-qubit reflectance histogram of the
    even (purple) and odd (orange) ground states.
    (h) Majorana-qubit reflectance histogram for
    the even (purple) and odd (orange) ground states of the Majorana qubit, with charge and amplifier noise.
    (i) Scatter plots showing the random reflectance
    values for the two Majorana-qubit basis 
    states.  
    Parameters:
    tunneling amplitudes for the charge qubit and spin qubit: $\Delta = 5\mu$eV;
    for the Majorana qubit:
    $\Delta_{\text{even}} = 10\mu$eV and  $\Delta_{\text{odd}} = 5\mu$eV;
    charge noise strength: $\sigma_\varepsilon = 4\, \mu$eV;
    amplifier noise strength: $\sigma_\Gamma = 0.02$;
    linear circuit response coefficient: $\alpha = 0.002$;
    The histogram bin width in top panels is $4 \times 10^{-4}$.
    Histograms a, b, d, e, g, h are based on 100,000 realizations.
    Scatter plots c, f, i are based on 5,000 realizations.
    \label{fig:chargenoise}
    }
    \end{figure*}

To illustrate the pdf of the measured reflectance,
we numerically generate random values for $\delta \varepsilon$,
and plot the resulting reflectance imaginary
part $\text{Im}(\Gamma)$
histograms
for the ground (purple) and the excited (orange)
state in Fig.~\ref{fig:chargenoise}a.
As expected, based on the detuning dependence
of the parametric capacitance (see 
Fig.~\ref{fig:preliminaries}b),
and the linear relation in Eq.~\eqref{eq:linearresponse},
both histograms show a sharp peak at finite
$\text{Im}(\Gamma)$, and a tail stretching toward zero.

 In this case, the maximum likelihood inference rule
implies a unit readout fidelity, $F=1$. 
This is because $\text{Im}(\Gamma) = 0$ is 
the maximum likelihood separator between the
two qubit states, and the ground (excited) state
always produces positive (negative) $\text{Im}(\Gamma)$,
hence it is impossible to make an erroneous state inference.

This perfect readout fidelity is degraded by 
amplifier noise, as illustrated in 
Fig.~\ref{eq:linearresponse}b.
To generate this histogram, we randomly sample both
charge noise and amplifier noise. 
That is, for a single realization $j$,
the detuning is 
$\varepsilon_{\text{R},j}$ and the reflectance is
$\Gamma_j = \Gamma(\varepsilon_{\text{R},j}) 
+ \delta \Gamma'_j
+ i \delta \Gamma''_j$,
where $\varepsilon_{\text{R},j}$ 
is drawn from a Gaussian distribution
with zero mean and standard deviation of $\sigma_\varepsilon$,
whereas $\delta \Gamma'$ and $\delta \Gamma''$ are both
drawn from a Gaussian distribution 
with zero mean and standard deviation of 
$\sigma_\Gamma$.

In Fig.~\ref{fig:chargenoise}b, we show a histogram
that is obtained by binning the data with respect to the
imaginary part of the reflectance, since in our model,
amplifier- and charge-noise-induced changes in the
real part do not affect the readout fidelity.
This histogram is essentially a smeared version of Fig.~\ref{fig:chargenoise}a,
and the smearing can be interpreted 
as a result of a convolution 
with the Gaussian representing amplifier noise.
To describe the picture expected for a large ensemble
of measurements, we show a scatter plot
of the randomly generated complex reflectance values
for the ground (blue) and excited (orange) 
states in Fig.~\ref{fig:chargenoise}c.

We can estimate the readout fidelity corresponding
to the maximum likelihood inference rule in
this random sampling framework. 
In short, we take the bins as in Fig.~\ref{fig:chargenoise}b,
and for each bin we assign an inferred state, e.g., 
the ground state if the bin's counts in the purple
histogram is greater than the bin's counts in the orange histogram.
This discretized maximum likelihood inference rule
defines our readout fidelity estimate, which, in the case
of the data in Fig.~\ref{fig:chargenoise}b,
yields $F \approx 0.98$.
The procedure for calculating $F$ is 
described in detail in appendix \ref{app:maximumlikelihood}.

To increase the readout fidelity in an experiment, 
reducing both 
the charge noise strength $\sigma_c$ 
and
the dimensionless amplifier noise strength
$\sigma_\Gamma$ 
seems to be a reasonable
strategy.
On the one hand, $\sigma_c$ is often an inherent,
uncontrollable  feature of the device.
On the other hand, there are two ways
to \emph{in situ} reduce 
$\sigma_\Gamma$, as suggested by Eq.~\eqref{eq:sigmagamma}.
The first way is 
to increase the integration 
time $t_\text{int}$.
This might, however, imply a reduced readout
fidelity due to inelastic
relaxation transitions between the qubit
basis states, if the integration time
exceeds the characteristic relaxation time scales.
(Positive consequences of 
charge relaxation for Majorana qubit readout are
discussed in section \ref{sec:discussion}.)
The second way to decrease the dimensionless
amplifier noise $\sigma_\Gamma$ is to increase
the amplitude of the ac probe voltage $V_\text{dev}$ at the 
gate. 
As we show below in sections \ref{sec:adiabatic}
and \ref{sec:transitions}, this increased
probe strength induces further significant 
changes beyond decreasing $\sigma_\Gamma$.

We wish to emphasize that the point clouds in
Fig.~\ref{fig:chargenoise}c
are anisotropic: they are elongated
along the imaginary axis.
This feature is reminiscent of quantum noise squeezing
\cite{ClerkRevModPhys2010}.
In our case, this anisotropy is a mundane consequence of 
charge noise, and has nothing to do with quantum-limited
amplification.
We also note that an interesting application of the
charge-noise-induced elongation of the reflectance
distributions can be to characterize the
strength $\sigma_\varepsilon$ of the charge noise. 

In case of the singlet-triplet 
spin qubit, we describe charge noise again
as a quasistatic random contribution 
(zero mean, standard deviation $\sigma_c$)
of the on-site energies of the two dots, 
which detunes the double dot from the
(1,1)-(0,2) crossing point.
Similarly to the charge-qubit case, 
this can be accounted for by assuming that
our detuning parameter $\varepsilon_\text{R}$
is a zero-mean 
Gaussian random variable with standard deviation
$\sigma_\varepsilon = \sqrt{2} \sigma_c$.
The reflectance histogram of the singlet state
is shown in
Fig.~\ref{fig:chargenoise}d.
The triplet histogram is not shown, 
since the reflectance is zero in that case,
irrespective of the random component of the detuning.
The singlet histogram has a shape
similar to that of the charge-qubit ground state in
Fig.~\ref{fig:chargenoise}a.
Amplifier noise causes a broadening of both peaks, 
as shown in the histograms
of the reflectance imaginary parts 
in Fig.~\ref{fig:chargenoise}e,
and the reflectance scatter plot in Fig.~\ref{fig:chargenoise}f.
The readout fidelity calculated from the histogram of
Fig.~\ref{fig:chargenoise}e is $F \approx 0.93$.

Finally, we illustrate the effect of charge noise 
on the readout of the Majorana qubit. 
Charge noise can be 
modelled as a random on-site energy on the readout dot
\cite{SzechenyiArxiv2019};
we denote the corresponding standard deviation 
by $\sigma_\varepsilon$. 
The resulting randomness of the reflectance for
the even and odd qubit states are 
shown in Fig.~\ref{fig:chargenoise}g. 
In this panel, amplifier noise is not accounted for. 
Remarkably, even charge noise alone makes readout imperfect
(in contrast to the charge qubit and spin qubits cases), 
since the orange and blue histograms in
Fig.~\ref{fig:chargenoise}g 
do overlap.
We estimate the readout fidelity as $F = 0.86$ from
the data in Fig.~\ref{fig:chargenoise}g.
Accounting for amplifier noise, the reflectance pdfs are
broadened, as shown in Fig.~\ref{fig:chargenoise}h, and 
fidelity is further reduced to 
$F \approx 0.74$.
A remarkable feature of the Majorana qubit result in 
Fig.~\ref{fig:chargenoise}h is
that the maximum inference rule does not yield a 
single threshold reflectance
separating the two qubit states, but two thresholds seperating 
three different regions: 
$\text{Im}(\Gamma) < 0.005$ (even),
$0.005< \text{Im}(\Gamma) < 0.02$ (odd), and
$0.02 < \text{Im}(\Gamma)$ (even).

\section{Overdrive effects in the adiabatic approximation}
\label{sec:adiabatic}

The results of the previous section suggest that by reducing
the dimensionless amplifier noise strength $\sigma_\Gamma$, 
the readout fidelity is increased. 
According to Eq.~\eqref{eq:sigmagamma},
one way to reduce the $\sigma_\Gamma$ is to increase the 
probe strength $V_\text{dev}$.
However, as we point out in this section 
and in the next section, 
the quantum dot in the device functions
as a nonlinear capacitor 
with a $V_\text{dev}$-dependent 
effective parametric capacitance
that decreases with increasing $V_\text{dev}$. 
For the charge and spin qubits, 
this behavior can lead to a saturation of the
signal-to-noise ratio and the readout fidelity
as $V_\text{dev}$ is increased. 
For the Majorana qubit, this behavior 
implies that there is an optimal $V_\text{dev}$ value, 
and increasing $V_\text{dev}$
beyond this optimum leads to a decreasing 
readout fidelity. %\cite{AhmedPhysRevApplied2018,SchaalPRL2020}.
These trends, which we call `overdrive effects',
can be seen already in the adiabatic 
approximation, 
when we assume that the qubit state follows the 
instantaneous
eigenstate of the driven qubit Hamiltonian. 
This section describes this adiabatic regime, whereas in 
the next one
we incorporate probe-induced transitions into the model. 

Let us take the charge qubit first, and comment on the spin 
and
Majorana qubits later.
For concreteness, we describe a scenario where the probe 
pulse 
is suddenly switched on at time $t=0$, is on for an 
integration
time $t_\text{int}$ that takes a few tens or hundreds of
probe periods, and then suddenly switched off. 
During this integration time, the double dot, tuned to the
tipping point $\varepsilon_\text{R} = 0$
as the readout position, 
and subject to the probe pulse, is described by the 
Hamiltonian 
    \begin{equation}
%    H(t)= \frac{\Delta }{2}\sigma _{x} + \frac{\varepsilon_R (t)}{2}\sigma _{z}
    H(t) = 
%    \frac{\Delta}{2}\sigma_x + \varepsilon_\text{R}(t) \frac{1-\sigma_z}{2}.
    \left(\begin{array}{cc}
    0 & \Delta/2 \\
    \Delta/2 & \varepsilon_\text{R}(t)
    \end{array} \right).
    \label{eq:reflectometryhamiltonian}
    \end{equation}
The time-dependent parameter $\varepsilon(t)$
represents the probe pulse:
    \begin{equation}
    \varepsilon_\text{R} (t)= eV_\text{dev}\sin \omega t
    \mbox{ (for $0<t<t_\text{int}$)}, 
    \label{eq:pulse}
    \end{equation}%
with amplitude $eV_\text{dev}$, and angular frequency 
$\omega$.
We assume that the integration time 
$t_\text{int}$ is much longer than the resonator transient
response (ring-up and ring-down times), 
such that the time dependence of $V_\text{dev}$ can be ignored. The impact of 
transients can be minimised by tailored amplitude modulation of the input signal \cite{WalterPRApp2017}.

In the results below, the frequency of the probe pulse
will be set to $\omega/(2\pi) = f = 325$ MHz.
Note that it corresponds to an energy quantum of 
$h f = \hbar \omega \approx 1.35 \mu\text{eV}$, 
and to a probe period $T = 1 /f \approx 3.07$ ns.

As long as probe-induced excitation processes
and inelastic relaxation processes can be neglected, 
a given energy eigenstate at $t=0$ will stay on its own dispersion branch for the whole integration time, 
i.e., the time evolution remains adiabatic. 
For example, the charge on the right dot 
in the adiabatically time-evolving charge-qubit ground
state is $-|e| n_\text{R}$, with 
\begin{equation}
    \label{eq:rightcharge}
    n_\text{R}(\varepsilon_\text{R}) = 
    \frac{1}{1+\left(\frac{\varepsilon_\text{R}}{\Delta}
    +\sqrt{1+\left(\frac{\varepsilon_\text{R}}{\Delta}\right)^2}\right)^2},
\end{equation}
whose the time dependence is described 
by substituting $\varepsilon_\text{R}(t)$ with 
Eq.~\eqref{eq:pulse}.
Together with the unit-lever-arm assumption (ii) 
in section \ref{sec:chargenoise}, 
this implies that time evolution of the 
charge on the gate-dot capacitor is
described by $|e| n_\text{R}(\varepsilon_\text{R}(t))$.

In turn, this implies that an effective parametric
capacitance can be attributed to 
this gate-dot capacitor, which is the ratio
of the in-phase Fourier component of the
charge response and the Fourier component of the
probing voltage, both taken at the probe frequency:
\begin{equation}
\label{eq:capgeneral}
    C_q =
    \frac{
    \frac{1}{t_\text{int}} \int_0^{t_\text{int}}
      dt |e| n_\text{R}(t) \sin(\omega t)
    }{
    \frac{1}{t_\text{int}}
    \int_0^{t_\text{int}}
      dt V_\text{dev} \sin (\omega t) \sin(\omega t)
    },
\end{equation}
The denominator can be readily evaluated, yielding
\begin{equation}
\label{eq:capadiabatic}
    C_q = \frac{2}{V_\text{dev}}
    \frac{1}{t_\text{int}} \int_0^{t_\text{int}}
      dt |e| n_\text{R}(t) \sin(\omega t).
\end{equation}
Furthermore, in the adiabatic approximation, the charge response 
$n_\text{R}(t)$ is periodic in time with period $T$, 
hence the formula
can be further simplifed via $t_\text{int} \mapsto T = 2\pi /\omega$:
\begin{equation}
    \label{eq:capforoneperiod}
    C_q = \frac{2}{V_\text{dev}}
    \frac{1}{T} \int_0^{T}
      dt |e| n_\text{R}(t) \sin(\omega t).
\end{equation}

In the adiabatic approximation, 
we obtain an analytical result for the parametric capacitance, 
by performing the integral of Eq.~\eqref{eq:capforoneperiod}
after inserting Eq.~\eqref{eq:rightcharge} and Eq.~\eqref{eq:pulse}:
\begin{equation}
\label{eq:ceff}
    C_q = \frac{2 |e|}{\pi V_\text{dev}} f_C(x),
\end{equation}
where 
the dimensionless function characterizing the 
capacitance is defined as 
\begin{equation}
\label{eq:fcfunction}
    f_C(x) = \frac{
    \left(1+x^2\right)
    E\left(\frac{x^2}{1+x^2}\right)
    -
    K\left(\frac{x^2}{1+x^2}\right)
    }{
    x \sqrt{1+x^2}
    },
\end{equation}
where
\begin{equation}
    x = \frac{|e| V_\text{dev}}{\Delta},
\end{equation}
furthermore, $E(x)$ is the 
complete elliptic integral of the second kind,
and $K(x)$ is the 
complete elliptic integral of the first kind.

\begin{figure}
    \begin{center}
        \includegraphics[width=1\columnwidth]{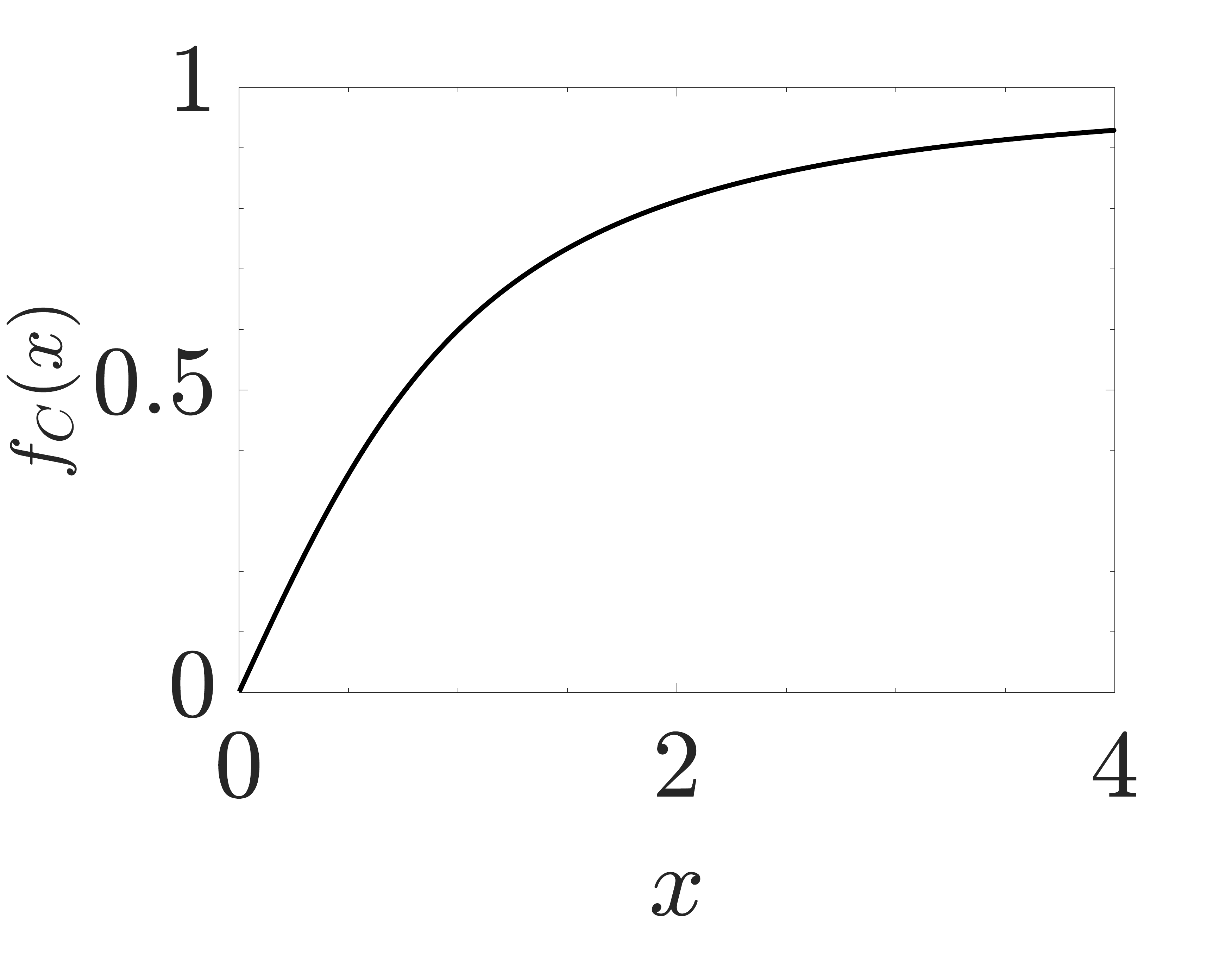}
    \end{center}
    \caption{The function $f_C(x)$ characterizing
    the probe-strength-dependence of the 
    effective capacitance, defined in Eq.~\eqref{eq:fcfunction}.
    \label{fig:fcfunction}}
    \end{figure}

The function $f_C(x)$
is shown in Fig.~\ref{fig:fcfunction}.
It is strictly monotonically increasing, and 
its asymptotics for small arguments is  
\begin{equation}
    \label{eq:smallarg}
    f_C(x \ll 1) \approx \frac{\pi}{4} x,
\end{equation}
whereas for large arguments it converges to 1:
\begin{equation}
    \label{eq:largearg}
    f_C(x \to \infty) = 1.
\end{equation}
Due to Eq.~\eqref{eq:smallarg}, the
weak-probe limit of $C_q$ in 
Eq.~\eqref{eq:ceff} is 
\begin{equation}
    C_q(|e| V_\text{dev} \ll \Delta) = 
    \frac{|e|^2}{2\Delta},
\end{equation}
as expected from the standard (linear) parametric
capacitance formula Eq.~\eqref{eq:chargequbitcapacitance}.

The analytical result for the probe-strength
dependence of the parametric capacitance
is shown in Fig.~\ref{fig:chargenoise}a,
where the solid purple (orange) line
corresponds to the ground (excited) state. 
The magnitude of the capacitance decreases with increasing
probe strength, which is the consequence of the
simple fact that the charge qubit is 
formed by a single electron, and 
therefore the charge on the right dot
is saturated for a large gate voltage. 

    \begin{figure*}
    \begin{center}
        \includegraphics[width=1.6\columnwidth]{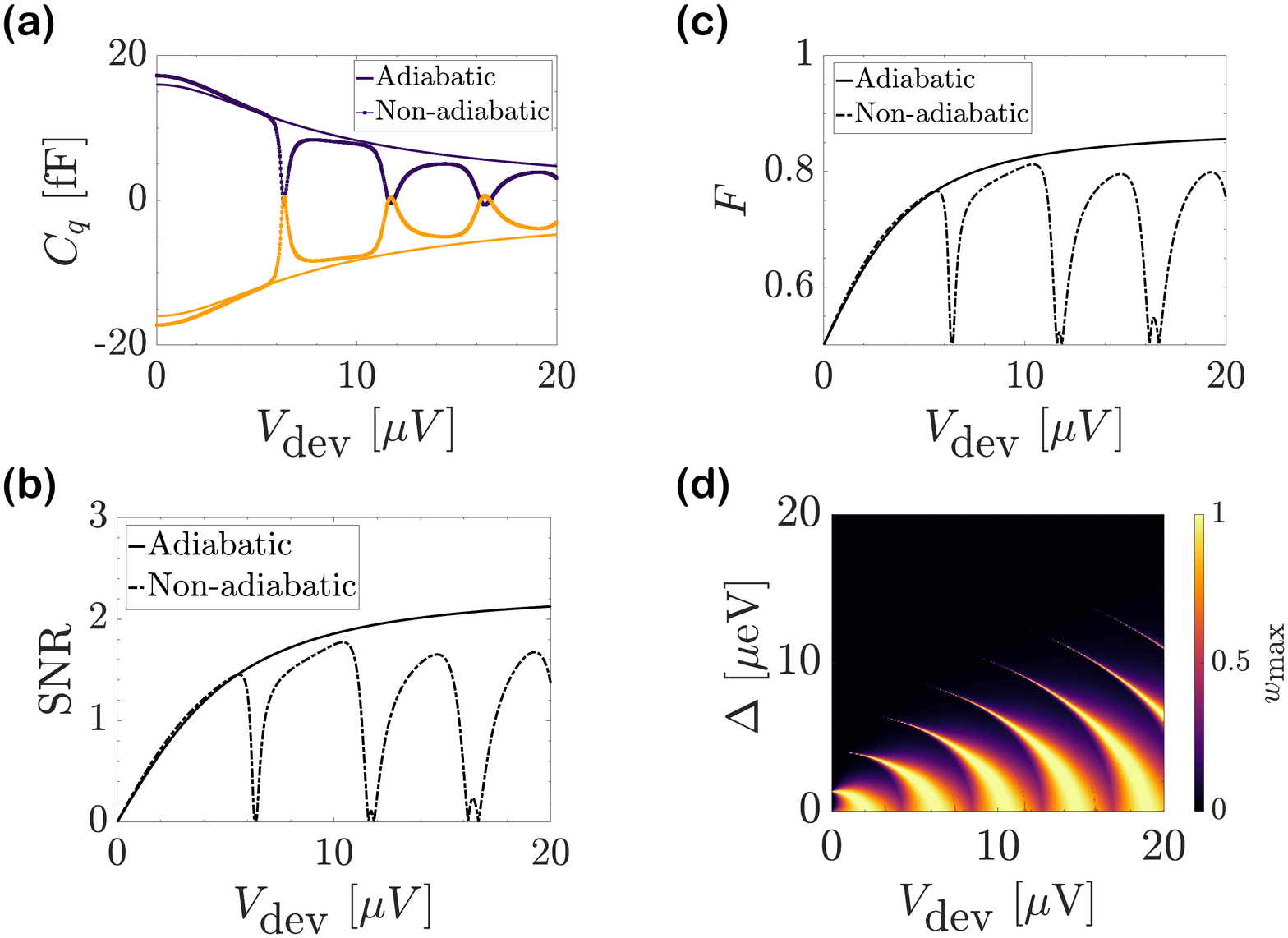}
    \end{center}
    \caption{
    Overdrive effects in reflectometry-based 
    readout of a charge qubit.
    (a) Probe-strength-dependent 
    effective parametric capacitance for
    ground state (purple) and excited state (orange).
    Solid: adiabatic approximation, Eq.~\eqref{eq:ceff}.
    Points: from solution of time-dependent
    Schr\"odinger equation.
    (b) Signal-to-noise ratio (SNR) 
    and 
    (c) readout fidelity 
    for a charge qubit,
    as function of probe strength,
    from adiabatic approximation (solid)
    and from time-dependent Schr\"odinger equation. 
    Both (b) and (c) shows a saturation 
    to a maximum as $V_\text{dev}$ increases.
    (d) Maximum transition probability $w_\text{max}$
    as function of probe strength $V_\text{dev}$ 
    and charge-qubit hybridization gap $\Delta$.
    Solid line shows the line of maximum SNR. 
    Parameters: tunnel splitting $\Delta = 5 \, \mu$eV,
    detuning $\varepsilon_\text{R} = 0$, 
    probe frequency: $f = 325$ MHz,
    integration time 
    $t_\text{int} = 325\, T = 1\, \mu$s.
    In (b), (c):
    noise temperature: $T_\text{n} = 0.1$ K,
    linear circuit response coefficient: 
    $\alpha = 0.01 \, \text{fF}^{-1}$,
    resistance $R=577\, \text{k}\Omega$.
        \label{fig:chargequbit}}
    \end{figure*}

Using \eqref{eq:ceff}, we now 
evaluate the SNR for the charge qubit via 
Eqs.~\eqref{eq:snrdef} and \eqref{eq:linearresponse}:
\begin{equation}
    \label{eq:snrchargequbit}
    \text{SNR} =
    \frac{
        4 \alpha |e| \sqrt{t_\text{int}}
    }{
        \pi\sqrt{k_B T_\text{n} R}
    } f_C(x).
\end{equation}
The dependence of the SNR on probe strength
$V_\text{dev}$ (via $x$) follows the increasing, but saturating,
dependence of $f_C(x)$ on $x$, as illustrated
in Fig.~\ref{fig:chargequbit}b,
where the solid line corresponds to the
adiabatic result \eqref{eq:snrchargequbit}. 
Due to Eq.~\eqref{eq:largearg}, we conclude
the best possible SNR is given by 
the prefactor of $f_C$ in Eq.~\eqref{eq:snrchargequbit}.
For the example parameter set
$\alpha = 0.01\, \text{fF}^{-1}$,
$t_\text{int} = 1 \, \mu$s,
$T_\text{n} = 0.1$ K,
and 
$R = 577\, \text{k}\Omega$, 
we estimate a maximal signal-to-noise ratio
$\text{SNR} \approx 2.29$,
as seen in Fig.~\ref{fig:chargequbit}b.
Correspondingly, the readout fidelity $F$
is also saturated, see solid line
in Fig.~\ref{fig:chargequbit}c.

The above paragraphs present a central result 
of this work: in contrast to the expectation
from Eq.~\eqref{eq:sigmagamma}, i.e., 
that the readout fidelity is increasing without
bounds as the probe strength $V_\text{dev}$ is increased, 
here we find that the readout fidelity is saturated.
The physical reason for this is that the charge response 
of the dot-gate capacitor is saturated for
a strong probe, since there
is only a single electron shuttled back and forth
in the charge qubit.

It is straightforward
to adapt the result \eqref{eq:snrchargequbit}
for the case of the singlet-triplet
spin qubit, yielding 
\begin{equation}
    \text{SNR} =
    \frac{
        2 \alpha |e| \sqrt{t_\text{int}}
    }{
        \pi\sqrt{k_B T_\text{n} R}
    } f_C(x').
\end{equation}
Note the factor of 2 difference
with respect to Eq.~\eqref{eq:snrchargequbit}, 
and the appearance of 
$x' = |e| V_\text{dev}/(2\sqrt{\Delta})$.
We conclude that the best achievable SNR for the
spin qubit is half of that for 
the charge qubit. 

\begin{figure}
    \begin{center}
        \includegraphics[width=1\columnwidth]{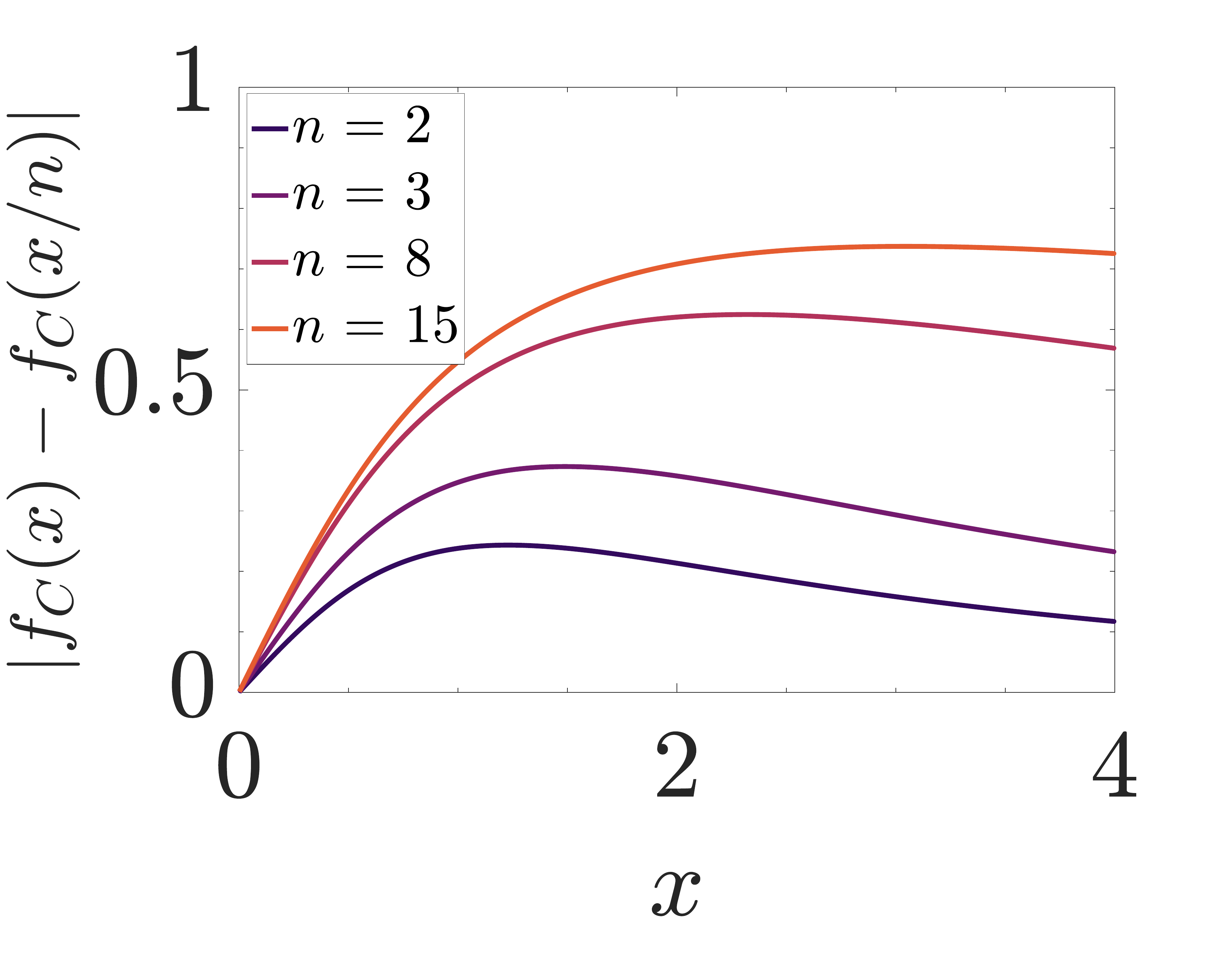}
    \end{center}
    \caption{Dimensionless function governing
    the signal-to-noise ratio 
    of Majorana-qubit readout as function
    of probe strength.
    Second term of the right hand side
    of Eq.~\eqref{eq:newsnrmajorana} is plotted, 
    for different values of 
    $n = \Delta_\text{even}/\Delta_\text{odd}$.
    Each curve has its unique maximum,
    corresponding to the best achievable signal-to-noise
    ratio according to Eq.~\eqref{eq:newsnrmajorana}.
    \label{fig:fcmajorana}}
    \end{figure}

A similar calculation can be carried
out for the Majorana qubit:
\begin{equation}
\label{eq:newsnrmajorana}
    \text{SNR} =
    \frac{
        2 \alpha |e| \sqrt{t_\text{int}}
    }{
        \pi\sqrt{k_B T_\text{n} R}
    } |f_C(x_\text{odd}) - f_C(x_\text{even})|,
\end{equation}
with 
$x_\text{odd} = |e| V_\text{dev}/\Delta_\text{odd}$
and
$x_\text{even} = |e| V_\text{dev}/\Delta_\text{even}$.
Since $f_C$ is bounded between 0 and 1
(see Fig.~\ref{fig:fcfunction}), 
therefore the
absolute value of the difference 
in Eq.~\eqref{eq:newsnrmajorana} is also
bounded between 0 and 1, 
as shown in Fig.~\ref{fig:fcmajorana}.
Therefore, we conclude that the best achievable
SNR for the Majorana-qubit readout
is given by the dimensionless fraction
in Eq.~\eqref{eq:newsnrmajorana}.

A key difference of the Majorana qubit 
compared to the charge and 
spin qubits is that for the former,
the SNR is opimtized by finding an appropriately
chosen finite probe strength, as illustrated in 
Fig.~\ref{fig:fcmajorana}.
For example, if $\Delta_\text{even} = 10\, \mu$eV
and $\Delta_\text{odd} = 5\, \mu$eV, 
and otherwise we set the parameter
values as specified in the caption 
of Fig.~\ref{fig:chargequbit},
then the dependence of the SNR on the
probe strength follows the blue curve
($n=2$) 
in Fig.~\ref{fig:fcmajorana}, 
implying that the best SNR is approximately
$0.25 \times 1.14 \approx 0.285$, 
reached for $x\approx 1.25$,
that is, a probe strength of
$V_\text{dev} \approx 1.25\ \Delta_\text{odd}/|e|
\approx 6.25 \, \mu$V.

In conclusion, we have shown that
for the charge and spin qubits, 
the adiabatic model considered
here predicts a saturation 
of the SNR and
readout fidelity as the probe strength is
increased;
whereas for the Majorana qubit, the same model
predicts the existence of an optimal
probe strength. 
We expect these effects to show up generically
in reflectometry-based qubit-readout
experiments, and to impose
a theoretical upper bound on the 
SNR and readout fidelity.
As we discuss in section \ref{sec:discussion}, 
this effect can also be used to calibrate
the amplitude $V_\text{dev}$
of the ac voltage that reaches the
gate electrode used for the reflectometry.

\section{Readout fidelity reduction due to
probe-induced transitions}
\label{sec:transitions}

In the previous section, we used the 
assumption of adiabaticity. 
In reality, the probe pulse is driving the
system, and thereby the latter can suffer 
transitions from one state to another,
and get driven out of the adiabatic regime. 
Such a transition can
decrease the contrast of the measurement.
Furthermore, this mechanism also induces
back-action on the measured system
by the measurement apparatus. 
Here, we focus on how the readout fidelity
defined above is degraded by these probe-induced
transitions.

For simplicity, we focus on the charge qubit, but
the effects described here apply also to the 
spin and Majorana qubits.
Here, we take the charge-qubit setup and
Hamiltonian 
as defined at the beginning of 
section \ref{sec:adiabatic}, but instead of
applying the adiabatic approximation, 
we describe the system by solving the 
time-dependent Schr\"odinger equation. 

Initially, the qubit is either
in the ground state or in the excited state
of the initial Hamiltonian 
$H(t=0) = \Delta \sigma_x / 2$,
that is, 
\begin{equation}
    \ket{\Psi(t=0)}  =
    \ket{\Psi_s} = \frac{1}{\sqrt{2}} \begin{bmatrix}
    1 \\ -\sigma_s
    \end{bmatrix}, \,  s \in \{g,e\}.
    \label{eq:initial}
\end{equation}
Recall the earlier definitions
$\sigma_g = +1$ and $\sigma_e = -1$.
The state evolves according to the time-dependent
Schr\"odinger equation governed by 
the Hamiltonian \eqref{eq:reflectometryhamiltonian};
we solve the Schr\"odinger equation numerically. 
This provides us with the time-evolving wave function
$\ket{\Psi(t)}$, 
and hence the 
time-evolving occupation of the right dot:
\begin{equation}
n_\text{R}(t) = \bra{\Psi(t)} \frac{1-\sigma_z}{2}
\ket{\Psi(t)}.   
\end{equation}
Note that this function depends on the
initial state being $g$ or $e$.
From the occupation, 
we calculate the probe-strength-dependent
effective parametric capacitances
$C_{q,g}$ and $C_{q,e}$ 
via Eq.~\eqref{eq:capadiabatic}, by
performing the integral numerically.
Both for the numerical solution of the 
Schr\"odinger equation and for the
numerical integration, we use a
time step $\Delta t = 0.001 \, T$.
Finally, we determine the reflectances $\Gamma_g$
and $\Gamma_e$ from
the effective parametric
capacitances via 
the linear-response condition Eq.~\eqref{eq:linearresponse},
and the SNR from Eq.~\eqref{eq:snrdef}.

We illustrate similarities and differences of this model,
as compared to the 
adiabatic model of section \ref{sec:adiabatic}, 
in Fig.~\ref{fig:chargequbit}.
Fig.~\ref{fig:chargequbit}a shows the 
effective parametric capacitances 
for the initial state $g$ 
(purple points) and $e$ (orange points), 
as obtained from the numerical solution of the
Schr\"odinger equation.
The data shows the same decreasing trend with increasing
probe strength $V_\text{dev}$ as in the 
adiabatic case, as revealed by comparing it to the
solid lines,  which are the analytical 
results.
However, there are sharp resonant dips 
in Fig.~\ref{fig:chargequbit}a
at well-defined
probe strength values, e.g., at 
$V_\text{dev} \approx 6.5 \, \mu$V, 
which are not there in the adiabatic result.
These resonant features are inherited by the
SNR (Fig.~\ref{fig:chargequbit}b) 
and by the readout fidelity 
(Fig.~\ref{fig:chargequbit}c).
In fact, around those special, resonant values of the probe 
strength, readout is imprecise.

The reason for those resonant dips in the 
parametric capacitance, SNR and readout fidelity is
the occurence of efficient transitions between
the two eigenstates of the static Hamiltonian. 
At these resonances, the dynamics of the qubit between
its instantaneous ground and excited states resembles
complete Rabi oscillations, which leads to a cancellation 
among the parametric 
capacitances of the ground and excited states
which have the same absolute value but 
different signs.

To support this explanation, we consider $g$
as the initial state, and 
calculate 
the maximal time-dependent transition probability 
over the duration of the integration time:
\begin{equation}
    w_\text{max} = \max_{t \in [0,t_\text{int}]} 
    \left|\braket{\Phi_e(t)}{\Psi(t)}\right|^2.
\end{equation}
Here, $\ket{\Phi_e(t)}$ is 
instantaneous excited state, that is, 
the positive-energy eigenstate 
of $H(t)$ of Eq.~\eqref{eq:reflectometryhamiltonian}.
The maximal transition probability $w_\text{max}$
is plotted 
in Fig.~\ref{fig:chargequbit}d 
as the function of the probe strength $V_\text{dev}$
and the tunnel splitting $\Delta$. 
This panel shows separate regions
of high transition probability that are bending
toward the $\Delta$ axis. 
These regions correspond to the the dips of
the SNR and readout fidelity. 
Each of these regions is related to a multi-photon 
transition: each of the regions approaches 
the $\Delta$ axis forming narrow `needles', 
whose endpoints extrapolated
to the $\Delta$ axis are 
$\approx 1.35\, \mu\text{eV}$,
$4.05 \, \mu\text{eV}$,
$6.75 \mu\text{eV}$, ...
accurately matching the series 
$h f$, $3hf$, $5 h f$, ..., 
where $f$ is the probe frequency.
Hence, these regions correspond to probe-induced
multi-photon transitions with odd photon number.
Alternatively, the high-transition-probability
regions can also be interpreted as
multi-passage Landau-Zener transitions \cite{ShevchenkoReview}.

Why are the even-photon transitions and the
corresponding features missing from
Fig.~\ref{fig:chargenoise}a-d? 
This is due the special multi-photon selection rules 
that arise because the qubit is tuned
exactly to the tipping point $\varepsilon_\text{R} = 0$
in Fig.~\ref{fig:chargenoise}. 
Then, the probe-induced driving is described
by a Hamiltonian that is transversal ($\propto \sigma_z$)
to the static Hamiltonian ($\propto \sigma_x$).
This restricts the active transitions to 
odd-photon ones\cite{ShirleyPhysRev1965,RomhanyiPRB2015}.

A further feature of Fig.~\ref{fig:chargequbit}a 
is that the adiabatic result and the
numerical result do not match exactly at
low probe strength $V_\text{dev} \to 0$. 
The reason is simple: in the numerical simulation, 
the sinusoidal probe pulse is switched on abruptly, without
a smooth envelope, hence the dynamics is not
perfectly adiabatic. 
Note that we use this model due to its simplicity:
in an experiment, the finite ring-up time of the
resonator will ensure a smooth switch-on of the
probe pulse and hence mitigate non-adiabatic features.

Above, we have described the probe-induced resonant 
dips in the readout fidelity, seen in Fig.~\ref{fig:chargequbit}c. 
Then it is natural to ask
how these features change in the presence of charge noise?
They do change, in two respects (not shown): 
(i) charge noise leads to the broadening
of the resonant dips, and (ii) it also leads to the 
appearance of extra dips. 

Feature (i) is rather natural: quasistatic charge noise
introduces a small randomness in the qubit splitting, 
and hence it slightly shifts the resonance positions. 
Averaging with respect to charge-noise configurations
hence leads to a broadening of the resonant dips. 

Feature (ii) is explained as
a consequence of the fact that the selection rule
for the probe-induced multi-photon transitions is
markedly changed when charge noise detunes the readout 
point from the tipping point.
Then, the transversality condition described above 
does not hold anymore, 
therefore even-photon transitions are also activated.

\section{Discussion}
\label{sec:discussion}
    
(1) \emph{Effects of relaxation processes.}
We comment on the validity of these
results in the presence of relaxation 
processes \cite{MizutaPRB2017,EsterliAPL2019,DAnjouPRB2019}.
For the charge qubit, our results hold
if the qubit relaxation time is longer than the 
integration time, but the results are not valid
if qubit relaxation is fast.
For example, in the zero-temperature limit, 
if the excited state relaxes during the integration 
time of the measurement, then
the contrast between the two reflectance values
is reduced.
In case of the spin qubit, our results are 
adequate in the limit when uphill charge relaxation
(from the singlet bonding state $S_g$ to the singlet 
antibonding state)
and spin relaxation take longer than the integration time. 

In case of the Majorana qubit, reflectometry-based
readout utilizes 
the ground states corresponding to the
two different fermion parities. 
Therefore, energetically downhill charge 
relaxation is actually useful, not
harmful \cite{KarzigPRB2017,KnappPRB2018,MunkArxiv2020}.
We expect that charge relaxation
enforces a behavior similar to the adiabatic
limit, but being a dissipative process, it should
add a resistive contribution (Sisyphus resistance) to
the response of the device \cite{EsterliAPL2019}, 
which could be quantified by calculating
the out-of-phase charge response to the ac
probe signal, that is, by replacing the 
sine in the numerator of Eq.~\eqref{eq:capgeneral} by 
a cosine.
In the presence of fast quasiparticle poisoning, 
when the parity of the coupled qubit-dot system switches
due to unwanted electron exchange with the environment faster
than the integration time,  our description
loses validity.
Note that quasiparticle poisoning is expected to be slow
in Majorana-qubit devices based on Coulomb-blockaded
islands
\cite{KarzigPRB2017,PluggeNewJPhys2017}.    
    
(2) 
\emph{Role of the lever arms.}
In a typical quantum-dot device, each quantum dot
is capacitively coupled to multiple metallic electrodes
(gates, contacts). 
For example, in a double-dot charge qubit, the right dot
is coupled to the plunger gate of the 
left dot as well. 
We have disregarded such cross-couplings by assumption
(ii) in section \ref{sec:chargequbit}.
It is straightforward to go beyond this
simplification, following, e.g., 
Ref.~\cite{MizutaPRB2017}.
We leave that for future work, but quote
one result, corresponding to 
a special case.
When the right dot couples dominantly to 
its gate electrode and its source 
and drain electrodes, but its capacitive coupling
to the left dot is negligible (see Fig.~1a of 
Ref.~\cite{MizutaPRB2017})
then the right dot -- gate coupling can be characterized
by the lever arm $0<\kappa<1$, and
the parametric capacitance calculated in the adiabatic approximation
[Eq.~\eqref{eq:ceff}] is modified as
\begin{equation}
\label{eq:ceffkappa}
    C_q = \frac{2 |e| \kappa}{\pi V_\text{dev}} f_C(\kappa x).
\end{equation}

(3) 
\emph{Further optimization of the signal-to-noise ratio.}
Throughout this work, we assumed that the
probe frequency is fixed at the 
eigenfrequency of the circuit in the absence
of parametric capacitance,  
and that in this working point, 
the circuit reflectance is a linear function 
of the qubit-state-dependent 
parametric capacitance [see Eq.~\eqref{eq:linearresponse}].
These assumptions are convenient, 
because they imply that the 
readout fidelity does not depend on the
absolute values of the qubit-state dependent 
effective capacitances, only on the 
difference of the two capacitance values.
Going beyond our approximations, 
as described in Appendix \ref{app:circuit}, 
a natural question arises:
how to select the probe frequency 
to achieve an optimal signal-to-noise ratio
for qubit readout?
A natural choice could be to tune the probe frequency
to the eigenfrequency of the circuit 
whose parametric capacitance is the average of
the two values corresponding to the two 
qubit states. 
Since the presence of charge noise distorts the
reflectance histograms from Gaussian, we expect
that finding the optimal probe frequency
is a nontrivial task in general.

    % \begin{figure}
    % \begin{center}
    %     \includegraphics[width=1\columnwidth]{Fidelity-ch-am.pdf}
    % \end{center}
    % \caption{The calculated readout fidelity as a function of $V_{\text{dev}}$ for the case of charge qubit while the both amplifier and charge noise is present when the tunnelling amplitude is  set $\Delta = 5\mu$eV.
    % \label{fig:Fidelity-ch-am}}
    % \end{figure}
    
\section{Conclusions}
\label{sec:conclusions}

We have provided a theoretical description of 
error mechanisms that degrade
qubit readout based on gate reflectometry:
slow charge noise and overdrive effects.
On-device charge noise creates a broadened distribution of 
the parametric capacitance of the device, and thereby
adds uncertainty to the measured reflectance, in addition to amplifier noise.
Hence charge noise leads to a reduced readout fidelity.
Overdriving, that is, applying a too strong probe pulse, can 
lead
to a scenario in which the charge response
of the quantum dot, and hence
the signal-to-noise ratio and the readout
fidelity, is saturated as the probe 
strength is increased. 
Remarkably, in the Majorana-qubit setting, 
we find that there is an optimal probe strength
leading to a maximal readout fidelity.
Overdriving can also cause coherent transitions between 
quantum states, 
which leads to further reduction of the readout fidelity
in the form of resonant dips.
We expect that our results can be conveniently applied for 
interpretation and
design of gate-reflectometry experiments. 
Furthermore, we think that the effects explored 
here can be utilized for device characterization: 
the  anisotropy of the reflectance histograms can be used
to infer the charge noise strength, whereas the 
overdrive effects could 
be used to calibrate the ac probe voltage amplitude that 
reaches the gate electrode. 
    
\acknowledgments
We acknowledge 
J. Asb\'oth, 
G. Burkard, G.~F\"ul\"op and M. Kocsis for useful discussions. 
This work was supported by the National Research Development and 
Innovation Office of Hungary within the Quantum Technology National 
Excellence Program (Project No. 2017-1.2.1-NKP-2017-00001), under OTKA 
Grants 124723, 132146 by the New National Excellence Program of the 
Ministry of Human Capacities,
and by the BME-Nanotechnology FIKP grant (BME FIKP-NAT).
MFGZ acknowledges support from the Royal Society.

\appendix
    
\section{An example circuit}
\label{app:circuit}

Here, we introduce an example 
circuit
to illustrate the key concepts 
that influence the physics discussed
in the main text. 
The circuit is shown in Fig.~\ref{fig:preliminaries}a.
For concreteness, we specify the circuit 
elements as 
$L = 405$ nH, 
$R = 576.6$ k$\Omega$, 
$C_c = 90$ fF and 
$C_0 = 486$ fF.
These values are similar to those in the experiment
of Ref.~\onlinecite{AhmedPhysRevApplied2018}, 
slightly modified to obtain perfect impedance matching:
at its resonance frequency $f=329.5$ MHz, 
the circuit impedance in the absence
of a parametric capacitance ($C_q=0$) reads as
\begin{equation}
\label{eq:deviceimpedance}
Z(\omega) = \left( R^{-1}+i \left(C_0 \omega -(L \omega )^{-1}\right)\right)^{-1} + \left(i C_c \omega \right)^{-1}.
\end{equation}
This impedance
is matched to a transmission-line characteristic
impedance of $Z_0 = 50\, \Omega$.
Note that perfect impedance matching
is not required for reflectometry, we choose it only
to simplify the discussion. 

\begin{figure}
   	\centering
   	\includegraphics[width=1\columnwidth]{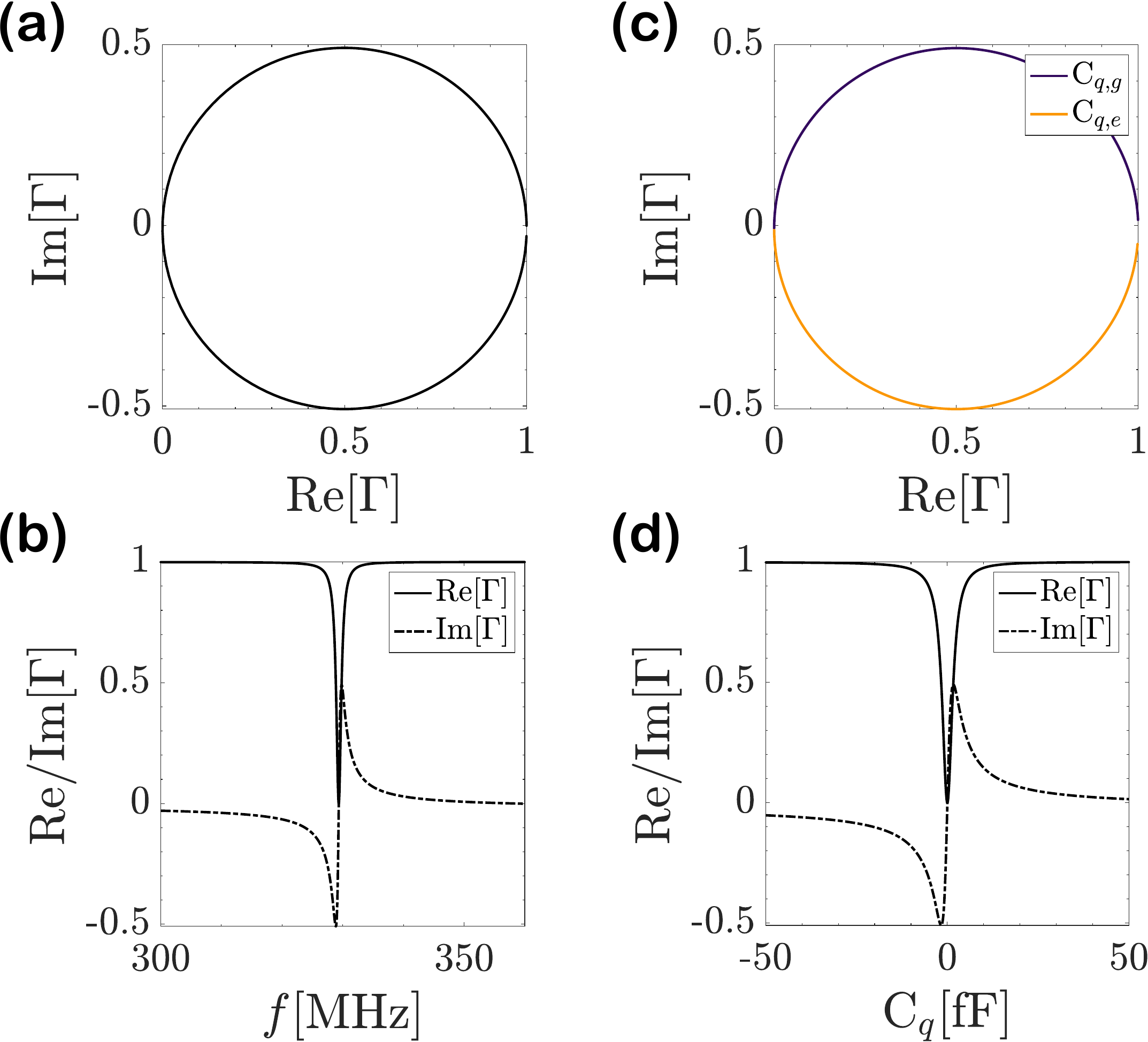}
   	\caption{
   	Circuit reflectance as a function of probe frequency 
   	and parametric capacitance. 
   	Reflectance of the circuit shown in Fig.~\ref{fig:preliminaries}a
   	as function of probe frequency in the absence of 
   	parametric capacitance, represented on 
   	the complex reflectance plane (a) and as a real an imaginary part (b).
   	Reflectance of the same circuit as function of 
   	parametric capacitance, 
   	at resonant probe frequency $f = 329.5$ MHz, 
   	represented on the complex reflectance plane (c)
   	and as a real and imaginary part (d).
   	See text of Appendix \ref{app:circuit} for circuit parameters.
   		\label{fig:linear-gamma}}
   \end{figure}   

The complex reflectance of the circuit is expressed as
\begin{equation}
\label{eq:gamma}
\Gamma =\frac{Z(\omega) - Z_0}{Z(\omega) + Z_0}.
\end{equation}
Real and imaginary parts of the reflectance
as the function of the probe
frequency $f=\omega/(2\pi)$ 
are shown in Fig.~\ref{fig:linear-gamma}b,
whereas the image of the reflectance map
$f\mapsto \Gamma$ is shown in Fig.~\ref{fig:linear-gamma}a.

Qubit-state readout can be performed due to the
dependence of the circuit impedance
on the parametric capacitance $C_q$ of the
device containing the quantum dot(s).
The qubit-state-dependent 
parametric capacitance modifies the 
device impedance via the substitution 
$C_0 \mapsto C_0 + C_q$ in Eq.~\eqref{eq:deviceimpedance},
which in turn changes the reflectance via
Eq.~\eqref{eq:gamma}.
For concreteness, we fix the probe frequency 
to the resonance circuit frequency in the absence
of parametric capacitance, $f = 329.5$ MHz, 
and show the dependence of the reflectance on
the parametric capacitance in Fig.~\ref{fig:linear-gamma}d.
The image of the reflectance map $C_q \mapsto \Gamma$ is
shown in Fig.~\ref{fig:linear-gamma}c.
The setting $C_q = 0$ corresponds to the meeting
point of the purple and orange arcs at $\Gamma = 0$, 
and for small but finite $C_q$, the reflectance change is 
linear in $C_q$ and happens along the $\text{Im}(\Gamma)$
axis. 
For this particular circuit, a linear expansion
in $C_q$ around zero yields
 \begin{equation}
     \Gamma \approx i\, 0.6 \, \text{fF}^{-1} C_q,
     \label{eq:series}
 \end{equation}
exemplifying the general linear-response
formula \eqref{eq:linearresponse} of the main text.
As illustrated by Fig.~\ref{fig:linear-gamma}c, this linear
relation is only approximate, and deviations from it can significantly 
change the results; in the main text we treat only this linear regime
for simplicity, and postpone to study the role of the deviations for
future work.

% \section{Amplifier noise}
 
% \noteandras{If we want to justify our $\sigma_\Gamma$ formula, that 
% could go here.}

\section{Maximum likelihood inference rule and readout fidelity}
\label{app:maximumlikelihood}

In Section \ref{sec:chargenoise}, we calculated the
readout fidelities (see $F = ...$ 
labels on Fig.~\ref{fig:chargenoise}a, b, d, e, g, h)
of charge, spin and Majorana qubits 
from the reflectance histograms shown in Fig.~\ref{fig:chargenoise}.
Here, we outline the procedure of that calculation. 

Consider, for example, the histogram shown in Fig.~\ref{fig:chargenoise}b.
Here, reflectance values $\Gamma_j$ from $N_r = 100,000$ realizations 
in the range $-0.05 < \text{Im}(\Gamma) < 0.05$ are
collected in $N_\text{bin} = 250$ bins, each bin with width 
$w_\text{bin} = 0.0004$.
For the charge qubit ground (excited) state, 
the number of reflectance values in bin $i \in \{1,\dots N_\text{bin}\}$
is denoted as $n_i^{(g)}$ ($n_i^{(e)}$).

With this notation, 
the maximum likelihood inference method is formalized as follows.
When the experimenter observes a reflectance value 
$\text{Im}(\Gamma)$ (it is sufficient to consider the imaginary part here),
she picks the bin $i$ that contains that reflectance value, 
compares $n_i^{(g)}$ and $n_i^{(e)}$, and
infers that the qubit was in state $g$ if 
$n_i^{(g)} > n_i^{(e)}$ and in state $e$ otherwise.
For later use, let us introduce the indicator function
of ground-state-dominated bins $\Theta(i)$, that is, 
$\Theta(i) = 1$ for those bins $i$
in which $n_i^{(g)} > n_i^{(e)}$ holds, 
and $\Theta(i) = 0$ for all other bin $i$.

Now, to characterize the accuracy of this maximum likelihood
inference procedure, we just count the number of incorrect
inferences, and convert that to a probability.
In particular, the probability of making an incorrect 
inference when the charge qubit is in state $g$
is
\begin{equation}
    p_g = \frac{1}{N_r}\sum_{i=1}^{N_\text{bin}} 
        (1-\Theta(i)) n_i^{(g)},
\end{equation}
whereas the probability of making an 
incorrect inference when the charge qubit is in state $e$
is
\begin{equation}
    p_e = \frac{1}{N_r}
    \sum_{i=1}^{N_\text{bin}} 
        \Theta(i) n_i^{(e)},
\end{equation}
Assuming a balanced a priori probability for the two qubit states
before the measurement, we conclude that the readout fidelity
is characterized by 
\begin{equation}
    F = 1-\frac{p_g + p_e}{2}.
\end{equation}

    \bibliography{OverdriveManuscript}

\begin{thebibliography}{47}%
\makeatletter
\providecommand \@ifxundefined [1]{%
 \@ifx{#1\undefined}
}%
\providecommand \@ifnum [1]{%
 \ifnum #1\expandafter \@firstoftwo
 \else \expandafter \@secondoftwo
 \fi
}%
\providecommand \@ifx [1]{%
 \ifx #1\expandafter \@firstoftwo
 \else \expandafter \@secondoftwo
 \fi
}%
\providecommand \natexlab [1]{#1}%
\providecommand \enquote  [1]{``#1''}%
\providecommand \bibnamefont  [1]{#1}%
\providecommand \bibfnamefont [1]{#1}%
\providecommand \citenamefont [1]{#1}%
\providecommand \href@noop [0]{\@secondoftwo}%
\providecommand \href [0]{\begingroup \@sanitize@url \@href}%
\providecommand \@href[1]{\@@startlink{#1}\@@href}%
\providecommand \@@href[1]{\endgroup#1\@@endlink}%
\providecommand \@sanitize@url [0]{\catcode `\\12\catcode `\$12\catcode
  `\&12\catcode `\#12\catcode `\^12\catcode `\_12\catcode `\%12\relax}%
\providecommand \@@startlink[1]{}%
\providecommand \@@endlink[0]{}%
\providecommand \url  [0]{\begingroup\@sanitize@url \@url }%
\providecommand \@url [1]{\endgroup\@href {#1}{\urlprefix }}%
\providecommand \urlprefix  [0]{URL }%
\providecommand \Eprint [0]{\href }%
\providecommand \doibase [0]{http://dx.doi.org/}%
\providecommand \selectlanguage [0]{\@gobble}%
\providecommand \bibinfo  [0]{\@secondoftwo}%
\providecommand \bibfield  [0]{\@secondoftwo}%
\providecommand \translation [1]{[#1]}%
\providecommand \BibitemOpen [0]{}%
\providecommand \bibitemStop [0]{}%
\providecommand \bibitemNoStop [0]{.\EOS\space}%
\providecommand \EOS [0]{\spacefactor3000\relax}%
\providecommand \BibitemShut  [1]{\csname bibitem#1\endcsname}%
\let\auto@bib@innerbib\@empty
%</preamble>
\bibitem [{\citenamefont {Hanson}\ \emph {et~al.}(2007)\citenamefont {Hanson},
  \citenamefont {Kouwenhoven}, \citenamefont {Petta}, \citenamefont {Tarucha},\
  and\ \citenamefont {Vandersypen}}]{HansonRevModPhys2007}%
  \BibitemOpen
  \bibfield  {author} {\bibinfo {author} {\bibfnamefont {R}~\bibnamefont
  {Hanson}}, \bibinfo {author} {\bibfnamefont {L~P}\ \bibnamefont
  {Kouwenhoven}}, \bibinfo {author} {\bibfnamefont {J~R}\ \bibnamefont
  {Petta}}, \bibinfo {author} {\bibfnamefont {S}~\bibnamefont {Tarucha}}, \
  and\ \bibinfo {author} {\bibfnamefont {L~M~K}\ \bibnamefont {Vandersypen}},\
  }\bibfield  {title} {\enquote {\bibinfo {title} {{Spins in few-electron
  quantum dots}},}\ }\href {\doibase 10.1103/RevModPhys.79.1217} {\bibfield
  {journal} {\bibinfo  {journal} {Reviews of Modern Physics}\ }\textbf
  {\bibinfo {volume} {79}},\ \bibinfo {pages} {1217--1265} (\bibinfo {year}
  {2007})}\BibitemShut {NoStop}%
\bibitem [{\citenamefont {Karzig}\ \emph {et~al.}(2017)\citenamefont {Karzig},
  \citenamefont {Knapp}, \citenamefont {Lutchyn}, \citenamefont {Bonderson},
  \citenamefont {Hastings}, \citenamefont {Nayak}, \citenamefont {Alicea},
  \citenamefont {Flensberg}, \citenamefont {Plugge}, \citenamefont {Oreg},
  \citenamefont {Marcus},\ and\ \citenamefont {Freedman}}]{KarzigPRB2017}%
  \BibitemOpen
  \bibfield  {author} {\bibinfo {author} {\bibfnamefont {Torsten}\ \bibnamefont
  {Karzig}}, \bibinfo {author} {\bibfnamefont {Christina}\ \bibnamefont
  {Knapp}}, \bibinfo {author} {\bibfnamefont {Roman~M}\ \bibnamefont
  {Lutchyn}}, \bibinfo {author} {\bibfnamefont {Parsa}\ \bibnamefont
  {Bonderson}}, \bibinfo {author} {\bibfnamefont {Matthew~B}\ \bibnamefont
  {Hastings}}, \bibinfo {author} {\bibfnamefont {Chetan}\ \bibnamefont
  {Nayak}}, \bibinfo {author} {\bibfnamefont {Jason}\ \bibnamefont {Alicea}},
  \bibinfo {author} {\bibfnamefont {Karsten}\ \bibnamefont {Flensberg}},
  \bibinfo {author} {\bibfnamefont {Stephan}\ \bibnamefont {Plugge}}, \bibinfo
  {author} {\bibfnamefont {Yuval}\ \bibnamefont {Oreg}}, \bibinfo {author}
  {\bibfnamefont {Charles~M}\ \bibnamefont {Marcus}}, \ and\ \bibinfo {author}
  {\bibfnamefont {Michael~H}\ \bibnamefont {Freedman}},\ }\bibfield  {title}
  {\enquote {\bibinfo {title} {{Scalable designs for
  quasiparticle-poisoning-protected topological quantum computation with
  Majorana zero modes}},}\ }\href {\doibase 10.1103/PhysRevB.95.235305}
  {\bibfield  {journal} {\bibinfo  {journal} {Physical Review B}\ }\textbf
  {\bibinfo {volume} {95}},\ \bibinfo {pages} {235305} (\bibinfo {year}
  {2017})}\BibitemShut {NoStop}%
\bibitem [{\citenamefont {Plugge}\ \emph {et~al.}(2017)\citenamefont {Plugge},
  \citenamefont {Rasmussen}, \citenamefont {Egger},\ and\ \citenamefont
  {Flensberg}}]{PluggeNewJPhys2017}%
  \BibitemOpen
  \bibfield  {author} {\bibinfo {author} {\bibfnamefont {Stephan}\ \bibnamefont
  {Plugge}}, \bibinfo {author} {\bibfnamefont {Asbj{\o}rn}\ \bibnamefont
  {Rasmussen}}, \bibinfo {author} {\bibfnamefont {Reinhold}\ \bibnamefont
  {Egger}}, \ and\ \bibinfo {author} {\bibfnamefont {Karsten}\ \bibnamefont
  {Flensberg}},\ }\bibfield  {title} {\enquote {\bibinfo {title} {{Majorana box
  qubits}},}\ }\href {\doibase 10.1088/1367-2630/aa54e1} {\bibfield  {journal}
  {\bibinfo  {journal} {New Journal of Physics}\ }\textbf {\bibinfo {volume}
  {19}},\ \bibinfo {pages} {012001} (\bibinfo {year} {2017})},\ \Eprint
  {http://arxiv.org/abs/1609.01697} {1609.01697} \BibitemShut {NoStop}%
\bibitem [{\citenamefont {van Veen}\ \emph {et~al.}(2019)\citenamefont {van
  Veen}, \citenamefont {de~Jong}, \citenamefont {Han}, \citenamefont {Prosko},
  \citenamefont {Krogstrup}, \citenamefont {Watson}, \citenamefont
  {Kouwenhoven},\ and\ \citenamefont {Pfaff}}]{vanVeenPRB2019}%
  \BibitemOpen
  \bibfield  {author} {\bibinfo {author} {\bibfnamefont {Jasper}\ \bibnamefont
  {van Veen}}, \bibinfo {author} {\bibfnamefont {Damaz}\ \bibnamefont
  {de~Jong}}, \bibinfo {author} {\bibfnamefont {Lin}\ \bibnamefont {Han}},
  \bibinfo {author} {\bibfnamefont {Christian}\ \bibnamefont {Prosko}},
  \bibinfo {author} {\bibfnamefont {Peter}\ \bibnamefont {Krogstrup}}, \bibinfo
  {author} {\bibfnamefont {John~D}\ \bibnamefont {Watson}}, \bibinfo {author}
  {\bibfnamefont {Leo~P}\ \bibnamefont {Kouwenhoven}}, \ and\ \bibinfo {author}
  {\bibfnamefont {Wolfgang}\ \bibnamefont {Pfaff}},\ }\bibfield  {title}
  {\enquote {\bibinfo {title} {{Revealing charge-tunneling processes between a
  quantum dot and a superconducting island through gate sensing}},}\ }\href
  {\doibase 10.1103/PhysRevB.100.174508} {\bibfield  {journal} {\bibinfo
  {journal} {Phys. Rev. B}\ }\textbf {\bibinfo {volume} {100}},\ \bibinfo
  {pages} {174508} (\bibinfo {year} {2019})}\BibitemShut {NoStop}%
\bibitem [{\citenamefont {de~Jong}\ \emph {et~al.}(2019)\citenamefont
  {de~Jong}, \citenamefont {van Veen}, \citenamefont {Binci}, \citenamefont
  {Singh}, \citenamefont {Krogstrup}, \citenamefont {Kouwenhoven},
  \citenamefont {Pfaff},\ and\ \citenamefont {Watson}}]{deJongPRApplied2019}%
  \BibitemOpen
  \bibfield  {author} {\bibinfo {author} {\bibfnamefont {Damaz}\ \bibnamefont
  {de~Jong}}, \bibinfo {author} {\bibfnamefont {Jasper}\ \bibnamefont {van
  Veen}}, \bibinfo {author} {\bibfnamefont {Luca}\ \bibnamefont {Binci}},
  \bibinfo {author} {\bibfnamefont {Amrita}\ \bibnamefont {Singh}}, \bibinfo
  {author} {\bibfnamefont {Peter}\ \bibnamefont {Krogstrup}}, \bibinfo {author}
  {\bibfnamefont {Leo~P}\ \bibnamefont {Kouwenhoven}}, \bibinfo {author}
  {\bibfnamefont {Wolfgang}\ \bibnamefont {Pfaff}}, \ and\ \bibinfo {author}
  {\bibfnamefont {John~D}\ \bibnamefont {Watson}},\ }\bibfield  {title}
  {\enquote {\bibinfo {title} {{Rapid Detection of Coherent Tunneling in an
  \text{InAs} Nanowire Quantum Dot through Dispersive Gate Sensing}},}\ }\href
  {\doibase 10.1103/PhysRevApplied.11.044061} {\bibfield  {journal} {\bibinfo
  {journal} {Phys. Rev. Appl.}\ }\textbf {\bibinfo {volume} {11}},\ \bibinfo
  {pages} {044061} (\bibinfo {year} {2019})}\BibitemShut {NoStop}%
\bibitem [{\citenamefont {Razmadze}\ \emph {et~al.}(2019)\citenamefont
  {Razmadze}, \citenamefont {Sabonis}, \citenamefont {Malinowski},
  \citenamefont {M{\'{e}}nard}, \citenamefont {Pauka}, \citenamefont {Nguyen},
  \citenamefont {van Zanten}, \citenamefont {O′Farrell}, \citenamefont
  {Suter}, \citenamefont {Krogstrup}, \citenamefont {Kuemmeth},\ and\
  \citenamefont {Marcus}}]{RazmadzePRApplied2019}%
  \BibitemOpen
  \bibfield  {author} {\bibinfo {author} {\bibfnamefont {Davydas}\ \bibnamefont
  {Razmadze}}, \bibinfo {author} {\bibfnamefont {Deividas}\ \bibnamefont
  {Sabonis}}, \bibinfo {author} {\bibfnamefont {Filip~K}\ \bibnamefont
  {Malinowski}}, \bibinfo {author} {\bibfnamefont {Gerbold~C}\ \bibnamefont
  {M{\'{e}}nard}}, \bibinfo {author} {\bibfnamefont {Sebastian}\ \bibnamefont
  {Pauka}}, \bibinfo {author} {\bibfnamefont {Hung}\ \bibnamefont {Nguyen}},
  \bibinfo {author} {\bibfnamefont {David~M.T.}\ \bibnamefont {van Zanten}},
  \bibinfo {author} {\bibfnamefont {Eoin~C.T.}\ \bibnamefont {O′Farrell}},
  \bibinfo {author} {\bibfnamefont {Judith}\ \bibnamefont {Suter}}, \bibinfo
  {author} {\bibfnamefont {Peter}\ \bibnamefont {Krogstrup}}, \bibinfo {author}
  {\bibfnamefont {Ferdinand}\ \bibnamefont {Kuemmeth}}, \ and\ \bibinfo
  {author} {\bibfnamefont {Charles~M}\ \bibnamefont {Marcus}},\ }\bibfield
  {title} {\enquote {\bibinfo {title} {{Radio-Frequency Methods for
  Majorana-Based Quantum Devices: Fast Charge Sensing and Phase-Diagram
  Mapping}},}\ }\href {\doibase 10.1103/PhysRevApplied.11.064011} {\bibfield
  {journal} {\bibinfo  {journal} {Phys. Rev. Appl.}\ }\textbf {\bibinfo
  {volume} {11}},\ \bibinfo {pages} {064011} (\bibinfo {year}
  {2019})}\BibitemShut {NoStop}%
\bibitem [{\citenamefont {Sabonis}\ \emph {et~al.}(2019)\citenamefont
  {Sabonis}, \citenamefont {O'Farrell}, \citenamefont {Razmadze}, \citenamefont
  {van Zanten}, \citenamefont {Suter}, \citenamefont {Krogstrup},\ and\
  \citenamefont {Marcus}}]{SabonisAPL2019}%
  \BibitemOpen
  \bibfield  {author} {\bibinfo {author} {\bibfnamefont {Deividas}\
  \bibnamefont {Sabonis}}, \bibinfo {author} {\bibfnamefont {Eoin C~T}\
  \bibnamefont {O'Farrell}}, \bibinfo {author} {\bibfnamefont {Davydas}\
  \bibnamefont {Razmadze}}, \bibinfo {author} {\bibfnamefont {David M~T}\
  \bibnamefont {van Zanten}}, \bibinfo {author} {\bibfnamefont {Judith}\
  \bibnamefont {Suter}}, \bibinfo {author} {\bibfnamefont {Peter}\ \bibnamefont
  {Krogstrup}}, \ and\ \bibinfo {author} {\bibfnamefont {Charles~M}\
  \bibnamefont {Marcus}},\ }\bibfield  {title} {\enquote {\bibinfo {title}
  {{Dispersive sensing in hybrid InAs/Al nanowires}},}\ }\href {\doibase
  10.1063/1.5116377} {\bibfield  {journal} {\bibinfo  {journal} {Appl. Phys.
  Lett.}\ }\textbf {\bibinfo {volume} {115}},\ \bibinfo {pages} {102601}
  (\bibinfo {year} {2019})}\BibitemShut {NoStop}%
\bibitem [{\citenamefont {Mi}\ \emph {et~al.}(2017)\citenamefont {Mi},
  \citenamefont {Cady}, \citenamefont {Zajac}, \citenamefont {Deelman},\ and\
  \citenamefont {Petta}}]{XiaoMiScience2017}%
  \BibitemOpen
  \bibfield  {author} {\bibinfo {author} {\bibfnamefont {X}~\bibnamefont {Mi}},
  \bibinfo {author} {\bibfnamefont {J~V}\ \bibnamefont {Cady}}, \bibinfo
  {author} {\bibfnamefont {D~M}\ \bibnamefont {Zajac}}, \bibinfo {author}
  {\bibfnamefont {P~W}\ \bibnamefont {Deelman}}, \ and\ \bibinfo {author}
  {\bibfnamefont {J~R}\ \bibnamefont {Petta}},\ }\bibfield  {title} {\enquote
  {\bibinfo {title} {{Strong coupling of a single electron in silicon to a
  microwave photon}},}\ }\href {\doibase 10.1126/science.aal2469} {\bibfield
  {journal} {\bibinfo  {journal} {Science}\ }\textbf {\bibinfo {volume}
  {355}},\ \bibinfo {pages} {156--158} (\bibinfo {year} {2017})}\BibitemShut
  {NoStop}%
\bibitem [{\citenamefont {Scarlino}\ \emph {et~al.}(2019)\citenamefont
  {Scarlino}, \citenamefont {van Woerkom}, \citenamefont {Stockklauser},
  \citenamefont {Koski}, \citenamefont {Collodo}, \citenamefont {Gasparinetti},
  \citenamefont {Reichl}, \citenamefont {Wegscheider}, \citenamefont {Ihn},
  \citenamefont {Ensslin},\ and\ \citenamefont {Wallraff}}]{ScarlinoPRL2019}%
  \BibitemOpen
  \bibfield  {author} {\bibinfo {author} {\bibfnamefont {P}~\bibnamefont
  {Scarlino}}, \bibinfo {author} {\bibfnamefont {D.~J.}\ \bibnamefont {van
  Woerkom}}, \bibinfo {author} {\bibfnamefont {A}~\bibnamefont {Stockklauser}},
  \bibinfo {author} {\bibfnamefont {J.~V.}\ \bibnamefont {Koski}}, \bibinfo
  {author} {\bibfnamefont {M.~C.}\ \bibnamefont {Collodo}}, \bibinfo {author}
  {\bibfnamefont {S}~\bibnamefont {Gasparinetti}}, \bibinfo {author}
  {\bibfnamefont {C}~\bibnamefont {Reichl}}, \bibinfo {author} {\bibfnamefont
  {W}~\bibnamefont {Wegscheider}}, \bibinfo {author} {\bibfnamefont
  {T}~\bibnamefont {Ihn}}, \bibinfo {author} {\bibfnamefont {K}~\bibnamefont
  {Ensslin}}, \ and\ \bibinfo {author} {\bibfnamefont {A}~\bibnamefont
  {Wallraff}},\ }\bibfield  {title} {\enquote {\bibinfo {title} {{All-Microwave
  Control and Dispersive Readout of Gate-Defined Quantum Dot Qubits in Circuit
  Quantum Electrodynamics}},}\ }\href {\doibase 10.1103/PhysRevLett.122.206802}
  {\bibfield  {journal} {\bibinfo  {journal} {Physical Review Letters}\
  }\textbf {\bibinfo {volume} {122}},\ \bibinfo {pages} {206802} (\bibinfo
  {year} {2019})}\BibitemShut {NoStop}%
\bibitem [{\citenamefont {Mizuta}\ \emph {et~al.}(2017)\citenamefont {Mizuta},
  \citenamefont {Otxoa}, \citenamefont {Betz},\ and\ \citenamefont
  {Gonzalez-Zalba}}]{MizutaPRB2017}%
  \BibitemOpen
  \bibfield  {author} {\bibinfo {author} {\bibfnamefont {R}~\bibnamefont
  {Mizuta}}, \bibinfo {author} {\bibfnamefont {R~M}\ \bibnamefont {Otxoa}},
  \bibinfo {author} {\bibfnamefont {A~C}\ \bibnamefont {Betz}}, \ and\ \bibinfo
  {author} {\bibfnamefont {M~F}\ \bibnamefont {Gonzalez-Zalba}},\ }\bibfield
  {title} {\enquote {\bibinfo {title} {{Quantum and tunneling capacitance in
  charge and spin qubits}},}\ }\href {\doibase 10.1103/PhysRevB.95.045414}
  {\bibfield  {journal} {\bibinfo  {journal} {Physical Review B}\ }\textbf
  {\bibinfo {volume} {95}},\ \bibinfo {pages} {045414} (\bibinfo {year}
  {2017})}\BibitemShut {NoStop}%
\bibitem [{\citenamefont {Ahmed}\ \emph {et~al.}(2018)\citenamefont {Ahmed},
  \citenamefont {Haigh}, \citenamefont {Schaal}, \citenamefont {Barraud},
  \citenamefont {Zhu}, \citenamefont {Lee}, \citenamefont {Amado},
  \citenamefont {Robinson}, \citenamefont {Rossi}, \citenamefont {Morton},\
  and\ \citenamefont {Gonzalez-Zalba}}]{AhmedPhysRevApplied2018}%
  \BibitemOpen
  \bibfield  {author} {\bibinfo {author} {\bibfnamefont {Imtiaz}\ \bibnamefont
  {Ahmed}}, \bibinfo {author} {\bibfnamefont {James~A}\ \bibnamefont {Haigh}},
  \bibinfo {author} {\bibfnamefont {Simon}\ \bibnamefont {Schaal}}, \bibinfo
  {author} {\bibfnamefont {Sylvain}\ \bibnamefont {Barraud}}, \bibinfo {author}
  {\bibfnamefont {Yi}~\bibnamefont {Zhu}}, \bibinfo {author} {\bibfnamefont
  {Chang-min}\ \bibnamefont {Lee}}, \bibinfo {author} {\bibfnamefont {Mario}\
  \bibnamefont {Amado}}, \bibinfo {author} {\bibfnamefont {Jason W~A}\
  \bibnamefont {Robinson}}, \bibinfo {author} {\bibfnamefont {Alessandro}\
  \bibnamefont {Rossi}}, \bibinfo {author} {\bibfnamefont {John J~L}\
  \bibnamefont {Morton}}, \ and\ \bibinfo {author} {\bibfnamefont {M~Fernando}\
  \bibnamefont {Gonzalez-Zalba}},\ }\bibfield  {title} {\enquote {\bibinfo
  {title} {{Radio-Frequency Capacitive Gate-Based Sensing}},}\ }\href {\doibase
  10.1103/PhysRevApplied.10.014018} {\bibfield  {journal} {\bibinfo  {journal}
  {Physical Review Applied}\ }\textbf {\bibinfo {volume} {10}},\ \bibinfo
  {pages} {014018} (\bibinfo {year} {2018})}\BibitemShut {NoStop}%
\bibitem [{\citenamefont {Colless}\ \emph {et~al.}(2013)\citenamefont
  {Colless}, \citenamefont {Mahoney}, \citenamefont {Hornibrook}, \citenamefont
  {Doherty}, \citenamefont {Lu}, \citenamefont {Gossard},\ and\ \citenamefont
  {Reilly}}]{CollessPRL2013}%
  \BibitemOpen
  \bibfield  {author} {\bibinfo {author} {\bibfnamefont {J.~I.}\ \bibnamefont
  {Colless}}, \bibinfo {author} {\bibfnamefont {A.~C.}\ \bibnamefont
  {Mahoney}}, \bibinfo {author} {\bibfnamefont {J.~M.}\ \bibnamefont
  {Hornibrook}}, \bibinfo {author} {\bibfnamefont {A.~C.}\ \bibnamefont
  {Doherty}}, \bibinfo {author} {\bibfnamefont {H.}~\bibnamefont {Lu}},
  \bibinfo {author} {\bibfnamefont {A.~C.}\ \bibnamefont {Gossard}}, \ and\
  \bibinfo {author} {\bibfnamefont {D.~J.}\ \bibnamefont {Reilly}},\ }\bibfield
   {title} {\enquote {\bibinfo {title} {{Dispersive Readout of a Few-Electron
  Double Quantum Dot with Fast rf Gate Sensors}},}\ }\href {\doibase
  10.1103/PhysRevLett.110.046805} {\bibfield  {journal} {\bibinfo  {journal}
  {Phys. Rev. Lett.}\ }\textbf {\bibinfo {volume} {110}},\ \bibinfo {pages}
  {046805} (\bibinfo {year} {2013})},\ \Eprint {http://arxiv.org/abs/1210.4645}
  {1210.4645} \BibitemShut {NoStop}%
\bibitem [{\citenamefont {Gonzalez-Zalba}\ \emph {et~al.}(2016)\citenamefont
  {Gonzalez-Zalba}, \citenamefont {Shevchenko}, \citenamefont {Barraud},
  \citenamefont {Johansson}, \citenamefont {Ferguson}, \citenamefont {Nori},\
  and\ \citenamefont {Betz}}]{Gonzalez-Zalba2016}%
  \BibitemOpen
  \bibfield  {author} {\bibinfo {author} {\bibfnamefont {M.~Fernando}\
  \bibnamefont {Gonzalez-Zalba}}, \bibinfo {author} {\bibfnamefont {Sergey~N}\
  \bibnamefont {Shevchenko}}, \bibinfo {author} {\bibfnamefont {Sylvain}\
  \bibnamefont {Barraud}}, \bibinfo {author} {\bibfnamefont {J~Robert}\
  \bibnamefont {Johansson}}, \bibinfo {author} {\bibfnamefont {Andrew~J}\
  \bibnamefont {Ferguson}}, \bibinfo {author} {\bibfnamefont {Franco}\
  \bibnamefont {Nori}}, \ and\ \bibinfo {author} {\bibfnamefont {Andreas~C.}\
  \bibnamefont {Betz}},\ }\bibfield  {title} {\enquote {\bibinfo {title}
  {{Gate-Sensing Coherent Charge Oscillations in a Silicon Field-Effect
  Transistor}},}\ }\href {\doibase 10.1021/acs.nanolett.5b04356} {\bibfield
  {journal} {\bibinfo  {journal} {Nano Lett.}\ }\textbf {\bibinfo {volume}
  {16}},\ \bibinfo {pages} {1614--1619} (\bibinfo {year} {2016})},\ \Eprint
  {http://arxiv.org/abs/1602.06004} {1602.06004} \BibitemShut {NoStop}%
\bibitem [{\citenamefont {Stockklauser}\ \emph {et~al.}(2017)\citenamefont
  {Stockklauser}, \citenamefont {Scarlino}, \citenamefont {Koski},
  \citenamefont {Gasparinetti}, \citenamefont {Andersen}, \citenamefont
  {Reichl}, \citenamefont {Wegscheider}, \citenamefont {Ihn}, \citenamefont
  {Ensslin},\ and\ \citenamefont {Wallraff}}]{StockklauserPRX2017}%
  \BibitemOpen
  \bibfield  {author} {\bibinfo {author} {\bibfnamefont {A}~\bibnamefont
  {Stockklauser}}, \bibinfo {author} {\bibfnamefont {P}~\bibnamefont
  {Scarlino}}, \bibinfo {author} {\bibfnamefont {J.~V.}\ \bibnamefont {Koski}},
  \bibinfo {author} {\bibfnamefont {S}~\bibnamefont {Gasparinetti}}, \bibinfo
  {author} {\bibfnamefont {C.~K.}\ \bibnamefont {Andersen}}, \bibinfo {author}
  {\bibfnamefont {C}~\bibnamefont {Reichl}}, \bibinfo {author} {\bibfnamefont
  {W}~\bibnamefont {Wegscheider}}, \bibinfo {author} {\bibfnamefont
  {T}~\bibnamefont {Ihn}}, \bibinfo {author} {\bibfnamefont {K}~\bibnamefont
  {Ensslin}}, \ and\ \bibinfo {author} {\bibfnamefont {A}~\bibnamefont
  {Wallraff}},\ }\bibfield  {title} {\enquote {\bibinfo {title} {{Strong
  Coupling Cavity QED with Gate-Defined Double Quantum Dots Enabled by a High
  Impedance Resonator}},}\ }\href {\doibase 10.1103/PhysRevX.7.011030}
  {\bibfield  {journal} {\bibinfo  {journal} {Phys. Rev. X}\ }\textbf {\bibinfo
  {volume} {7}},\ \bibinfo {pages} {011030} (\bibinfo {year}
  {2017})}\BibitemShut {NoStop}%
\bibitem [{\citenamefont {Chatterjee}\ \emph {et~al.}(2018)\citenamefont
  {Chatterjee}, \citenamefont {Shevchenko}, \citenamefont {Barraud},
  \citenamefont {Otxoa}, \citenamefont {Nori}, \citenamefont {Morton},\ and\
  \citenamefont {Gonzalez-Zalba}}]{ChatterjeePRB2018}%
  \BibitemOpen
  \bibfield  {author} {\bibinfo {author} {\bibfnamefont {Anasua}\ \bibnamefont
  {Chatterjee}}, \bibinfo {author} {\bibfnamefont {Sergey~N}\ \bibnamefont
  {Shevchenko}}, \bibinfo {author} {\bibfnamefont {Sylvain}\ \bibnamefont
  {Barraud}}, \bibinfo {author} {\bibfnamefont {Rub{\'{e}}n~M}\ \bibnamefont
  {Otxoa}}, \bibinfo {author} {\bibfnamefont {Franco}\ \bibnamefont {Nori}},
  \bibinfo {author} {\bibfnamefont {John J~L}\ \bibnamefont {Morton}}, \ and\
  \bibinfo {author} {\bibfnamefont {M~Fernando}\ \bibnamefont
  {Gonzalez-Zalba}},\ }\bibfield  {title} {\enquote {\bibinfo {title} {{A
  silicon-based single-electron interferometer coupled to a fermionic sea}},}\
  }\href {\doibase 10.1103/PhysRevB.97.045405} {\bibfield  {journal} {\bibinfo
  {journal} {Phys. Rev. B}\ }\textbf {\bibinfo {volume} {97}},\ \bibinfo
  {pages} {045405} (\bibinfo {year} {2018})}\BibitemShut {NoStop}%
\bibitem [{\citenamefont {Stehlik}\ \emph {et~al.}(2015)\citenamefont
  {Stehlik}, \citenamefont {Liu}, \citenamefont {Quintana}, \citenamefont
  {Eichler}, \citenamefont {Hartke},\ and\ \citenamefont
  {Petta}}]{StehlikPRApp2015}%
  \BibitemOpen
  \bibfield  {author} {\bibinfo {author} {\bibfnamefont {J}~\bibnamefont
  {Stehlik}}, \bibinfo {author} {\bibfnamefont {Y.-Y.}\ \bibnamefont {Liu}},
  \bibinfo {author} {\bibfnamefont {C.~M.}\ \bibnamefont {Quintana}}, \bibinfo
  {author} {\bibfnamefont {C}~\bibnamefont {Eichler}}, \bibinfo {author}
  {\bibfnamefont {T.~R.}\ \bibnamefont {Hartke}}, \ and\ \bibinfo {author}
  {\bibfnamefont {J.~R.}\ \bibnamefont {Petta}},\ }\bibfield  {title} {\enquote
  {\bibinfo {title} {{Fast Charge Sensing of a Cavity-Coupled Double Quantum
  Dot Using a Josephson Parametric Amplifier}},}\ }\href {\doibase
  10.1103/PhysRevApplied.4.014018} {\bibfield  {journal} {\bibinfo  {journal}
  {Phys. Rev. Appl.}\ }\textbf {\bibinfo {volume} {4}},\ \bibinfo {pages}
  {014018} (\bibinfo {year} {2015})}\BibitemShut {NoStop}%
\bibitem [{\citenamefont {Schaal}\ \emph {et~al.}(2020)\citenamefont {Schaal},
  \citenamefont {Ahmed}, \citenamefont {Haigh}, \citenamefont {Hutin},
  \citenamefont {Bertrand}, \citenamefont {Barraud}, \citenamefont {Vinet},
  \citenamefont {Lee}, \citenamefont {Stelmashenko}, \citenamefont {Robinson},
  \citenamefont {Qiu}, \citenamefont {Hacohen-Gourgy}, \citenamefont {Siddiqi},
  \citenamefont {Gonzalez-Zalba},\ and\ \citenamefont
  {Morton}}]{SchaalPRL2020}%
  \BibitemOpen
  \bibfield  {author} {\bibinfo {author} {\bibfnamefont {S}~\bibnamefont
  {Schaal}}, \bibinfo {author} {\bibfnamefont {I}~\bibnamefont {Ahmed}},
  \bibinfo {author} {\bibfnamefont {J.~A.}\ \bibnamefont {Haigh}}, \bibinfo
  {author} {\bibfnamefont {L}~\bibnamefont {Hutin}}, \bibinfo {author}
  {\bibfnamefont {B}~\bibnamefont {Bertrand}}, \bibinfo {author} {\bibfnamefont
  {S}~\bibnamefont {Barraud}}, \bibinfo {author} {\bibfnamefont
  {M}~\bibnamefont {Vinet}}, \bibinfo {author} {\bibfnamefont {C.-M.}\
  \bibnamefont {Lee}}, \bibinfo {author} {\bibfnamefont {N}~\bibnamefont
  {Stelmashenko}}, \bibinfo {author} {\bibfnamefont {J.~W.~A.}\ \bibnamefont
  {Robinson}}, \bibinfo {author} {\bibfnamefont {J.~Y.}\ \bibnamefont {Qiu}},
  \bibinfo {author} {\bibfnamefont {S}~\bibnamefont {Hacohen-Gourgy}}, \bibinfo
  {author} {\bibfnamefont {I}~\bibnamefont {Siddiqi}}, \bibinfo {author}
  {\bibfnamefont {M.~F.}\ \bibnamefont {Gonzalez-Zalba}}, \ and\ \bibinfo
  {author} {\bibfnamefont {J.~J.~L.}\ \bibnamefont {Morton}},\ }\bibfield
  {title} {\enquote {\bibinfo {title} {{Fast Gate-Based Readout of Silicon
  Quantum Dots Using Josephson Parametric Amplification}},}\ }\href {\doibase
  10.1103/PhysRevLett.124.067701} {\bibfield  {journal} {\bibinfo  {journal}
  {Phys. Rev. Lett.}\ }\textbf {\bibinfo {volume} {124}},\ \bibinfo {pages}
  {067701} (\bibinfo {year} {2020})}\BibitemShut {NoStop}%
\bibitem [{\citenamefont {Petit}\ \emph {et~al.}(2018)\citenamefont {Petit},
  \citenamefont {Boter}, \citenamefont {Eenink}, \citenamefont {Droulers},
  \citenamefont {Tagliaferri}, \citenamefont {Li}, \citenamefont {Franke},
  \citenamefont {Singh}, \citenamefont {Clarke}, \citenamefont {Schouten},
  \citenamefont {Dobrovitski}, \citenamefont {Vandersypen},\ and\ \citenamefont
  {Veldhorst}}]{PetitPRL2019}%
  \BibitemOpen
  \bibfield  {author} {\bibinfo {author} {\bibfnamefont {L}~\bibnamefont
  {Petit}}, \bibinfo {author} {\bibfnamefont {J.~M.}\ \bibnamefont {Boter}},
  \bibinfo {author} {\bibfnamefont {H.~G.~J.}\ \bibnamefont {Eenink}}, \bibinfo
  {author} {\bibfnamefont {G}~\bibnamefont {Droulers}}, \bibinfo {author}
  {\bibfnamefont {M.~L.~V.}\ \bibnamefont {Tagliaferri}}, \bibinfo {author}
  {\bibfnamefont {R}~\bibnamefont {Li}}, \bibinfo {author} {\bibfnamefont
  {D.~P.}\ \bibnamefont {Franke}}, \bibinfo {author} {\bibfnamefont {K.~J.}\
  \bibnamefont {Singh}}, \bibinfo {author} {\bibfnamefont {J.~S.}\ \bibnamefont
  {Clarke}}, \bibinfo {author} {\bibfnamefont {R.~N.}\ \bibnamefont
  {Schouten}}, \bibinfo {author} {\bibfnamefont {V.~V.}\ \bibnamefont
  {Dobrovitski}}, \bibinfo {author} {\bibfnamefont {L.~M.~K.}\ \bibnamefont
  {Vandersypen}}, \ and\ \bibinfo {author} {\bibfnamefont {M}~\bibnamefont
  {Veldhorst}},\ }\bibfield  {title} {\enquote {\bibinfo {title} {{Spin
  Lifetime and Charge Noise in Hot Silicon Quantum Dot Qubits}},}\ }\href
  {\doibase 10.1103/PhysRevLett.121.076801} {\bibfield  {journal} {\bibinfo
  {journal} {Phys. Rev. Lett.}\ }\textbf {\bibinfo {volume} {121}},\ \bibinfo
  {pages} {076801} (\bibinfo {year} {2018})}\BibitemShut {NoStop}%
\bibitem [{\citenamefont {Petersson}\ \emph
  {et~al.}(2010{\natexlab{a}})\citenamefont {Petersson}, \citenamefont {Smith},
  \citenamefont {Anderson}, \citenamefont {Atkinson}, \citenamefont {Jones},\
  and\ \citenamefont {Ritchie}}]{Petersson2010aNanolett}%
  \BibitemOpen
  \bibfield  {author} {\bibinfo {author} {\bibfnamefont {K.~D.}\ \bibnamefont
  {Petersson}}, \bibinfo {author} {\bibfnamefont {C.~G.}\ \bibnamefont
  {Smith}}, \bibinfo {author} {\bibfnamefont {D.}~\bibnamefont {Anderson}},
  \bibinfo {author} {\bibfnamefont {P.}~\bibnamefont {Atkinson}}, \bibinfo
  {author} {\bibfnamefont {G.~A.C.}\ \bibnamefont {Jones}}, \ and\ \bibinfo
  {author} {\bibfnamefont {D.~A.}\ \bibnamefont {Ritchie}},\ }\bibfield
  {title} {\enquote {\bibinfo {title} {{Charge and spin state readout of a
  double quantum dot coupled to a resonator}},}\ }\href {\doibase
  10.1021/nl100663w} {\bibfield  {journal} {\bibinfo  {journal} {Nano Lett.}\
  }\textbf {\bibinfo {volume} {10}},\ \bibinfo {pages} {2789--2793} (\bibinfo
  {year} {2010}{\natexlab{a}})},\ \Eprint {http://arxiv.org/abs/1004.4047}
  {1004.4047} \BibitemShut {NoStop}%
\bibitem [{\citenamefont {M{\"{u}}ller}\ and\ \citenamefont
  {Krishnaswamy}(1974)}]{Muller1974}%
  \BibitemOpen
  \bibfield  {author} {\bibinfo {author} {\bibfnamefont {Erwin~W}\ \bibnamefont
  {M{\"{u}}ller}}\ and\ \bibinfo {author} {\bibfnamefont {S.V.}\ \bibnamefont
  {Krishnaswamy}},\ }\bibfield  {title} {\enquote {\bibinfo {title} {{Energy
  deficits in pulsed field evaporation and deficit compensated atom‐probe
  designs}},}\ }\href {\doibase 10.1063/1.1686808} {\bibfield  {journal}
  {\bibinfo  {journal} {Rev. Sci. Instrum.}\ }\textbf {\bibinfo {volume}
  {45}},\ \bibinfo {pages} {1053--1059} (\bibinfo {year} {1974})}\BibitemShut
  {NoStop}%
\bibitem [{\citenamefont {Clerk}\ \emph {et~al.}(2010)\citenamefont {Clerk},
  \citenamefont {Devoret}, \citenamefont {Girvin}, \citenamefont {Marquardt},\
  and\ \citenamefont {Schoelkopf}}]{ClerkRevModPhys2010}%
  \BibitemOpen
  \bibfield  {author} {\bibinfo {author} {\bibfnamefont {A~A}\ \bibnamefont
  {Clerk}}, \bibinfo {author} {\bibfnamefont {M~H}\ \bibnamefont {Devoret}},
  \bibinfo {author} {\bibfnamefont {S~M}\ \bibnamefont {Girvin}}, \bibinfo
  {author} {\bibfnamefont {Florian}\ \bibnamefont {Marquardt}}, \ and\ \bibinfo
  {author} {\bibfnamefont {R~J}\ \bibnamefont {Schoelkopf}},\ }\bibfield
  {title} {\enquote {\bibinfo {title} {{Introduction to quantum noise,
  measurement, and amplification}},}\ }\href {\doibase
  10.1103/RevModPhys.82.1155} {\bibfield  {journal} {\bibinfo  {journal} {Rev.
  Mod. Phys.}\ }\textbf {\bibinfo {volume} {82}},\ \bibinfo {pages}
  {1155--1208} (\bibinfo {year} {2010})}\BibitemShut {NoStop}%
\bibitem [{\citenamefont {Petta}\ \emph {et~al.}(2005)\citenamefont {Petta},
  \citenamefont {Johnson}, \citenamefont {Taylor}, \citenamefont {Laird},
  \citenamefont {Yacoby}, \citenamefont {Lukin}, \citenamefont {Marcus},
  \citenamefont {Hanson},\ and\ \citenamefont {Gossard}}]{PettaScience2005}%
  \BibitemOpen
  \bibfield  {author} {\bibinfo {author} {\bibfnamefont {J.~R.}\ \bibnamefont
  {Petta}}, \bibinfo {author} {\bibfnamefont {A.~C.}\ \bibnamefont {Johnson}},
  \bibinfo {author} {\bibfnamefont {J.~M.}\ \bibnamefont {Taylor}}, \bibinfo
  {author} {\bibfnamefont {E.~A.}\ \bibnamefont {Laird}}, \bibinfo {author}
  {\bibfnamefont {A.}~\bibnamefont {Yacoby}}, \bibinfo {author} {\bibfnamefont
  {M.~D.}\ \bibnamefont {Lukin}}, \bibinfo {author} {\bibfnamefont {C.~M.}\
  \bibnamefont {Marcus}}, \bibinfo {author} {\bibfnamefont {M.~P.}\
  \bibnamefont {Hanson}}, \ and\ \bibinfo {author} {\bibfnamefont {A.~C.}\
  \bibnamefont {Gossard}},\ }\bibfield  {title} {\enquote {\bibinfo {title}
  {Coherent manipulation of coupled electron spins in semiconductor quantum
  dots},}\ }\href {\doibase 10.1126/science.1116955} {\bibfield  {journal}
  {\bibinfo  {journal} {Science}\ }\textbf {\bibinfo {volume} {309}},\ \bibinfo
  {pages} {2180--2184} (\bibinfo {year} {2005})}\BibitemShut {NoStop}%
\bibitem [{\citenamefont {Kitaev}(2001)}]{Kitaev2001unpaired}%
  \BibitemOpen
  \bibfield  {author} {\bibinfo {author} {\bibfnamefont {A~Yu}\ \bibnamefont
  {Kitaev}},\ }\bibfield  {title} {\enquote {\bibinfo {title} {{Unpaired
  Majorana fermions in quantum wires}},}\ }\href {\doibase
  10.1070/1063-7869/44/10S/S29} {\bibfield  {journal} {\bibinfo  {journal}
  {Physics-Uspekhi}\ }\textbf {\bibinfo {volume} {44}},\ \bibinfo {pages}
  {131--136} (\bibinfo {year} {2001})}\BibitemShut {NoStop}%
\bibitem [{\citenamefont {Alicea}\ \emph {et~al.}(2011)\citenamefont {Alicea},
  \citenamefont {Oreg}, \citenamefont {Refael}, \citenamefont {von Oppen},\
  and\ \citenamefont {Fisher}}]{AliceaNatPhys2011}%
  \BibitemOpen
  \bibfield  {author} {\bibinfo {author} {\bibfnamefont {Jason}\ \bibnamefont
  {Alicea}}, \bibinfo {author} {\bibfnamefont {Yuval}\ \bibnamefont {Oreg}},
  \bibinfo {author} {\bibfnamefont {Gil}\ \bibnamefont {Refael}}, \bibinfo
  {author} {\bibfnamefont {Felix}\ \bibnamefont {von Oppen}}, \ and\ \bibinfo
  {author} {\bibfnamefont {Matthew P~A}\ \bibnamefont {Fisher}},\ }\bibfield
  {title} {\enquote {\bibinfo {title} {{Non-Abelian statistics and topological
  quantum information processing in 1D wire networks}},}\ }\href {\doibase
  10.1038/nphys1915} {\bibfield  {journal} {\bibinfo  {journal} {Nature
  Physics}\ }\textbf {\bibinfo {volume} {7}},\ \bibinfo {pages} {412--417}
  (\bibinfo {year} {2011})},\ \Eprint {http://arxiv.org/abs/1006.4395}
  {1006.4395} \BibitemShut {NoStop}%
\bibitem [{\citenamefont {Flensberg}(2011)}]{FlensbergPRL2011}%
  \BibitemOpen
  \bibfield  {author} {\bibinfo {author} {\bibfnamefont {Karsten}\ \bibnamefont
  {Flensberg}},\ }\bibfield  {title} {\enquote {\bibinfo {title} {{Non-Abelian
  Operations on Majorana Fermions via Single-Charge Control}},}\ }\href
  {\doibase 10.1103/PhysRevLett.106.090503} {\bibfield  {journal} {\bibinfo
  {journal} {Phys. Rev. Lett.}\ }\textbf {\bibinfo {volume} {106}},\ \bibinfo
  {pages} {090503} (\bibinfo {year} {2011})}\BibitemShut {NoStop}%
\bibitem [{\citenamefont {Gharavi}\ \emph {et~al.}(2016)\citenamefont
  {Gharavi}, \citenamefont {Hoving},\ and\ \citenamefont
  {Baugh}}]{GharaviPRB2016}%
  \BibitemOpen
  \bibfield  {author} {\bibinfo {author} {\bibfnamefont {Kaveh}\ \bibnamefont
  {Gharavi}}, \bibinfo {author} {\bibfnamefont {Darryl}\ \bibnamefont
  {Hoving}}, \ and\ \bibinfo {author} {\bibfnamefont {Jonathan}\ \bibnamefont
  {Baugh}},\ }\bibfield  {title} {\enquote {\bibinfo {title} {{Readout of
  Majorana parity states using a quantum dot}},}\ }\href {\doibase
  10.1103/PhysRevB.94.155417} {\bibfield  {journal} {\bibinfo  {journal} {Phys.
  Rev. B}\ }\textbf {\bibinfo {volume} {94}},\ \bibinfo {pages} {155417}
  (\bibinfo {year} {2016})}\BibitemShut {NoStop}%
\bibitem [{\citenamefont {Knapp}\ \emph {et~al.}(2018)\citenamefont {Knapp},
  \citenamefont {Karzig}, \citenamefont {Lutchyn},\ and\ \citenamefont
  {Nayak}}]{KnappPRB2018}%
  \BibitemOpen
  \bibfield  {author} {\bibinfo {author} {\bibfnamefont {Christina}\
  \bibnamefont {Knapp}}, \bibinfo {author} {\bibfnamefont {Torsten}\
  \bibnamefont {Karzig}}, \bibinfo {author} {\bibfnamefont {Roman~M}\
  \bibnamefont {Lutchyn}}, \ and\ \bibinfo {author} {\bibfnamefont {Chetan}\
  \bibnamefont {Nayak}},\ }\bibfield  {title} {\enquote {\bibinfo {title}
  {{Dephasing of Majorana-based qubits}},}\ }\href {\doibase
  10.1103/PhysRevB.97.125404} {\bibfield  {journal} {\bibinfo  {journal} {Phys.
  Rev. B}\ }\textbf {\bibinfo {volume} {97}},\ \bibinfo {pages} {125404}
  (\bibinfo {year} {2018})}\BibitemShut {NoStop}%
\bibitem [{\citenamefont {Munk}\ \emph {et~al.}(2020)\citenamefont {Munk},
  \citenamefont {Schulenborg}, \citenamefont {Egger},\ and\ \citenamefont
  {Flensberg}}]{MunkArxiv2020}%
  \BibitemOpen
  \bibfield  {author} {\bibinfo {author} {\bibfnamefont {Morten I.~K.}\
  \bibnamefont {Munk}}, \bibinfo {author} {\bibfnamefont {Jens}\ \bibnamefont
  {Schulenborg}}, \bibinfo {author} {\bibfnamefont {Reinhold}\ \bibnamefont
  {Egger}}, \ and\ \bibinfo {author} {\bibfnamefont {Karsten}\ \bibnamefont
  {Flensberg}},\ }\href@noop {} {\enquote {\bibinfo {title} {{Parity-to-charge
  conversion in Majorana qubit readout}},}\ } (\bibinfo {year} {2020}),\
  \Eprint {http://arxiv.org/abs/2004.02123} {arXiv:2004.02123
  [cond-mat.mes-hall]} \BibitemShut {NoStop}%
\bibitem [{\citenamefont {Steiner}\ and\ \citenamefont {von
  Oppen}(2020)}]{SteinerArxiv2020}%
  \BibitemOpen
  \bibfield  {author} {\bibinfo {author} {\bibfnamefont {J.~F.}\ \bibnamefont
  {Steiner}}\ and\ \bibinfo {author} {\bibfnamefont {F.}~\bibnamefont {von
  Oppen}},\ }\href@noop {} {\enquote {\bibinfo {title} {{Readout of Majorana
  qubits}},}\ } (\bibinfo {year} {2020}),\ \bibinfo {note}
  {arXiv:2004.02124}\BibitemShut {NoStop}%
\bibitem [{\citenamefont {Freeman}\ \emph {et~al.}(2016)\citenamefont
  {Freeman}, \citenamefont {Schoenfield},\ and\ \citenamefont
  {Jiang}}]{FreemanAPL2016}%
  \BibitemOpen
  \bibfield  {author} {\bibinfo {author} {\bibfnamefont {Blake~M}\ \bibnamefont
  {Freeman}}, \bibinfo {author} {\bibfnamefont {Joshua~S}\ \bibnamefont
  {Schoenfield}}, \ and\ \bibinfo {author} {\bibfnamefont {HongWen}\
  \bibnamefont {Jiang}},\ }\bibfield  {title} {\enquote {\bibinfo {title}
  {{Comparison of low frequency charge noise in identically patterned Si/SiO 2
  and Si/SiGe quantum dots}},}\ }\href {\doibase 10.1063/1.4954700} {\bibfield
  {journal} {\bibinfo  {journal} {Appl. Phys. Lett.}\ }\textbf {\bibinfo
  {volume} {108}},\ \bibinfo {pages} {253108} (\bibinfo {year}
  {2016})}\BibitemShut {NoStop}%
\bibitem [{\citenamefont {Yoneda}\ \emph {et~al.}(2018)\citenamefont {Yoneda},
  \citenamefont {Takeda}, \citenamefont {Otsuka}, \citenamefont {Nakajima},
  \citenamefont {Delbecq}, \citenamefont {Allison}, \citenamefont {Honda},
  \citenamefont {Kodera}, \citenamefont {Oda}, \citenamefont {Hoshi},
  \citenamefont {Usami}, \citenamefont {Itoh},\ and\ \citenamefont
  {Tarucha}}]{YonedaNatNano2018}%
  \BibitemOpen
  \bibfield  {author} {\bibinfo {author} {\bibfnamefont {Jun}\ \bibnamefont
  {Yoneda}}, \bibinfo {author} {\bibfnamefont {Kenta}\ \bibnamefont {Takeda}},
  \bibinfo {author} {\bibfnamefont {Tomohiro}\ \bibnamefont {Otsuka}}, \bibinfo
  {author} {\bibfnamefont {Takashi}\ \bibnamefont {Nakajima}}, \bibinfo
  {author} {\bibfnamefont {Matthieu~R.}\ \bibnamefont {Delbecq}}, \bibinfo
  {author} {\bibfnamefont {Giles}\ \bibnamefont {Allison}}, \bibinfo {author}
  {\bibfnamefont {Takumu}\ \bibnamefont {Honda}}, \bibinfo {author}
  {\bibfnamefont {Tetsuo}\ \bibnamefont {Kodera}}, \bibinfo {author}
  {\bibfnamefont {Shunri}\ \bibnamefont {Oda}}, \bibinfo {author}
  {\bibfnamefont {Yusuke}\ \bibnamefont {Hoshi}}, \bibinfo {author}
  {\bibfnamefont {Noritaka}\ \bibnamefont {Usami}}, \bibinfo {author}
  {\bibfnamefont {Kohei~M.}\ \bibnamefont {Itoh}}, \ and\ \bibinfo {author}
  {\bibfnamefont {Seigo}\ \bibnamefont {Tarucha}},\ }\bibfield  {title}
  {\enquote {\bibinfo {title} {{A quantum-dot spin qubit with coherence limited
  by charge noise and fidelity higher than 99.9{\%}}},}\ }\href {\doibase
  10.1038/s41565-017-0014-x} {\bibfield  {journal} {\bibinfo  {journal} {Nat.
  Nanotechnol.}\ }\textbf {\bibinfo {volume} {13}},\ \bibinfo {pages}
  {102--106} (\bibinfo {year} {2018})}\BibitemShut {NoStop}%
\bibitem [{\citenamefont {Boter}\ \emph {et~al.}(2019)\citenamefont {Boter},
  \citenamefont {Xue}, \citenamefont {Krähenmann}, \citenamefont {Watson},
  \citenamefont {Premakumar}, \citenamefont {Ward}, \citenamefont {Savage},
  \citenamefont {Lagally}, \citenamefont {Friesen}, \citenamefont
  {Coppersmith}, \citenamefont {Eriksson}, \citenamefont {Joynt},\ and\
  \citenamefont {Vandersypen}}]{BoterArxiv2019}%
  \BibitemOpen
  \bibfield  {author} {\bibinfo {author} {\bibfnamefont {Jelmer~M.}\
  \bibnamefont {Boter}}, \bibinfo {author} {\bibfnamefont {Xiao}\ \bibnamefont
  {Xue}}, \bibinfo {author} {\bibfnamefont {Tobias~S.}\ \bibnamefont
  {Krähenmann}}, \bibinfo {author} {\bibfnamefont {Thomas~F.}\ \bibnamefont
  {Watson}}, \bibinfo {author} {\bibfnamefont {Vickram~N.}\ \bibnamefont
  {Premakumar}}, \bibinfo {author} {\bibfnamefont {Daniel~R.}\ \bibnamefont
  {Ward}}, \bibinfo {author} {\bibfnamefont {Donald~E.}\ \bibnamefont
  {Savage}}, \bibinfo {author} {\bibfnamefont {Max~G.}\ \bibnamefont
  {Lagally}}, \bibinfo {author} {\bibfnamefont {Mark}\ \bibnamefont {Friesen}},
  \bibinfo {author} {\bibfnamefont {Susan~N.}\ \bibnamefont {Coppersmith}},
  \bibinfo {author} {\bibfnamefont {Mark~A.}\ \bibnamefont {Eriksson}},
  \bibinfo {author} {\bibfnamefont {Robert}\ \bibnamefont {Joynt}}, \ and\
  \bibinfo {author} {\bibfnamefont {Lieven M.~K.}\ \bibnamefont
  {Vandersypen}},\ }\href@noop {} {\enquote {\bibinfo {title} {{Spatial Noise
  Correlations in a Si/SiGe Two-Qubit Device from Bell State Coherences}},}\ }
  (\bibinfo {year} {2019}),\ \Eprint {http://arxiv.org/abs/1906.02731}
  {arXiv:1906.02731 [cond-mat.mes-hall]} \BibitemShut {NoStop}%
\bibitem [{\citenamefont {Wang}\ \emph {et~al.}(2012)\citenamefont {Wang},
  \citenamefont {Bishop}, \citenamefont {Kestner}, \citenamefont {Barnes},
  \citenamefont {Sun},\ and\ \citenamefont {{Das Sarma}}}]{XinWang2012NatComm}%
  \BibitemOpen
  \bibfield  {author} {\bibinfo {author} {\bibfnamefont {Xin}\ \bibnamefont
  {Wang}}, \bibinfo {author} {\bibfnamefont {Lev~S}\ \bibnamefont {Bishop}},
  \bibinfo {author} {\bibfnamefont {J.P.}\ \bibnamefont {Kestner}}, \bibinfo
  {author} {\bibfnamefont {Edwin}\ \bibnamefont {Barnes}}, \bibinfo {author}
  {\bibfnamefont {Kai}\ \bibnamefont {Sun}}, \ and\ \bibinfo {author}
  {\bibfnamefont {S}~\bibnamefont {{Das Sarma}}},\ }\bibfield  {title}
  {\enquote {\bibinfo {title} {{Composite pulses for robust universal control
  of singlet–triplet qubits}},}\ }\href {\doibase 10.1038/ncomms2003}
  {\bibfield  {journal} {\bibinfo  {journal} {Nat. Commun.}\ }\textbf {\bibinfo
  {volume} {3}},\ \bibinfo {pages} {997} (\bibinfo {year} {2012})}\BibitemShut
  {NoStop}%
\bibitem [{\citenamefont {Tosi}\ \emph {et~al.}(2017)\citenamefont {Tosi},
  \citenamefont {Mohiyaddin}, \citenamefont {Schmitt}, \citenamefont {Tenberg},
  \citenamefont {Rahman}, \citenamefont {Klimeck},\ and\ \citenamefont
  {Morello}}]{TosiNatComm2017}%
  \BibitemOpen
  \bibfield  {author} {\bibinfo {author} {\bibfnamefont {Guilherme}\
  \bibnamefont {Tosi}}, \bibinfo {author} {\bibfnamefont {Fahd~A.}\
  \bibnamefont {Mohiyaddin}}, \bibinfo {author} {\bibfnamefont {Vivien}\
  \bibnamefont {Schmitt}}, \bibinfo {author} {\bibfnamefont {Stefanie}\
  \bibnamefont {Tenberg}}, \bibinfo {author} {\bibfnamefont {Rajib}\
  \bibnamefont {Rahman}}, \bibinfo {author} {\bibfnamefont {Gerhard}\
  \bibnamefont {Klimeck}}, \ and\ \bibinfo {author} {\bibfnamefont {Andrea}\
  \bibnamefont {Morello}},\ }\bibfield  {title} {\enquote {\bibinfo {title}
  {Silicon quantum processor with robust long-distance qubit couplings},}\
  }\href@noop {} {\bibfield  {journal} {\bibinfo  {journal} {Nature
  Communications}\ }\textbf {\bibinfo {volume} {8}},\ \bibinfo {pages} {450}
  (\bibinfo {year} {2017})}\BibitemShut {NoStop}%
\bibitem [{\citenamefont {Watson}\ \emph {et~al.}(2018)\citenamefont {Watson},
  \citenamefont {Philips}, \citenamefont {Kawakami}, \citenamefont {Ward},
  \citenamefont {Scarlino}, \citenamefont {Veldhorst}, \citenamefont {Savage},
  \citenamefont {Lagally}, \citenamefont {Friesen}, \citenamefont
  {Coppersmith}, \citenamefont {Eriksson},\ and\ \citenamefont
  {Vandersypen}}]{WatsonNature2018}%
  \BibitemOpen
  \bibfield  {author} {\bibinfo {author} {\bibfnamefont {T~F}\ \bibnamefont
  {Watson}}, \bibinfo {author} {\bibfnamefont {S~G~J}\ \bibnamefont {Philips}},
  \bibinfo {author} {\bibfnamefont {E}~\bibnamefont {Kawakami}}, \bibinfo
  {author} {\bibfnamefont {D~R}\ \bibnamefont {Ward}}, \bibinfo {author}
  {\bibfnamefont {P}~\bibnamefont {Scarlino}}, \bibinfo {author} {\bibfnamefont
  {M}~\bibnamefont {Veldhorst}}, \bibinfo {author} {\bibfnamefont {D~E}\
  \bibnamefont {Savage}}, \bibinfo {author} {\bibfnamefont {M~G}\ \bibnamefont
  {Lagally}}, \bibinfo {author} {\bibfnamefont {Mark}\ \bibnamefont {Friesen}},
  \bibinfo {author} {\bibfnamefont {S~N}\ \bibnamefont {Coppersmith}}, \bibinfo
  {author} {\bibfnamefont {M~A}\ \bibnamefont {Eriksson}}, \ and\ \bibinfo
  {author} {\bibfnamefont {L~M~K}\ \bibnamefont {Vandersypen}},\ }\bibfield
  {title} {\enquote {\bibinfo {title} {{A programmable two-qubit quantum
  processor in silicon}},}\ }\href {\doibase 10.1038/nature25766} {\bibfield
  {journal} {\bibinfo  {journal} {Nature}\ }\textbf {\bibinfo {volume} {555}},\
  \bibinfo {pages} {633--637} (\bibinfo {year} {2018})}\BibitemShut {NoStop}%
\bibitem [{\citenamefont {Dovzhenko}\ \emph {et~al.}(2011)\citenamefont
  {Dovzhenko}, \citenamefont {Stehlik}, \citenamefont {Petersson},
  \citenamefont {Petta}, \citenamefont {Lu},\ and\ \citenamefont
  {Gossard}}]{DovzhenkoPRB2011}%
  \BibitemOpen
  \bibfield  {author} {\bibinfo {author} {\bibfnamefont {Y}~\bibnamefont
  {Dovzhenko}}, \bibinfo {author} {\bibfnamefont {J}~\bibnamefont {Stehlik}},
  \bibinfo {author} {\bibfnamefont {K~D}\ \bibnamefont {Petersson}}, \bibinfo
  {author} {\bibfnamefont {J~R}\ \bibnamefont {Petta}}, \bibinfo {author}
  {\bibfnamefont {H}~\bibnamefont {Lu}}, \ and\ \bibinfo {author}
  {\bibfnamefont {A~C}\ \bibnamefont {Gossard}},\ }\bibfield  {title} {\enquote
  {\bibinfo {title} {{Nonadiabatic quantum control of a semiconductor charge
  qubit}},}\ }\href {\doibase 10.1103/PhysRevB.84.161302} {\bibfield  {journal}
  {\bibinfo  {journal} {Phys. Rev. B}\ }\textbf {\bibinfo {volume} {84}},\
  \bibinfo {pages} {161302} (\bibinfo {year} {2011})}\BibitemShut {NoStop}%
\bibitem [{\citenamefont {Petta}\ \emph {et~al.}(2004)\citenamefont {Petta},
  \citenamefont {Johnson}, \citenamefont {Marcus}, \citenamefont {Hanson},\
  and\ \citenamefont {Gossard}}]{PettaPRL2004}%
  \BibitemOpen
  \bibfield  {author} {\bibinfo {author} {\bibfnamefont {J~R}\ \bibnamefont
  {Petta}}, \bibinfo {author} {\bibfnamefont {A~C}\ \bibnamefont {Johnson}},
  \bibinfo {author} {\bibfnamefont {C~M}\ \bibnamefont {Marcus}}, \bibinfo
  {author} {\bibfnamefont {M~P}\ \bibnamefont {Hanson}}, \ and\ \bibinfo
  {author} {\bibfnamefont {A~C}\ \bibnamefont {Gossard}},\ }\bibfield  {title}
  {\enquote {\bibinfo {title} {{Manipulation of a Single Charge in a Double
  Quantum Dot}},}\ }\href {\doibase 10.1103/PhysRevLett.93.186802} {\bibfield
  {journal} {\bibinfo  {journal} {Phys. Rev. Lett.}\ }\textbf {\bibinfo
  {volume} {93}},\ \bibinfo {pages} {186802} (\bibinfo {year}
  {2004})}\BibitemShut {NoStop}%
\bibitem [{\citenamefont {Petersson}\ \emph
  {et~al.}(2010{\natexlab{b}})\citenamefont {Petersson}, \citenamefont {Petta},
  \citenamefont {Lu},\ and\ \citenamefont {Gossard}}]{PeterssonPRL2010}%
  \BibitemOpen
  \bibfield  {author} {\bibinfo {author} {\bibfnamefont {K~D}\ \bibnamefont
  {Petersson}}, \bibinfo {author} {\bibfnamefont {J~R}\ \bibnamefont {Petta}},
  \bibinfo {author} {\bibfnamefont {H}~\bibnamefont {Lu}}, \ and\ \bibinfo
  {author} {\bibfnamefont {A~C}\ \bibnamefont {Gossard}},\ }\bibfield  {title}
  {\enquote {\bibinfo {title} {{Quantum Coherence in a One-Electron
  Semiconductor Charge Qubit}},}\ }\href {\doibase
  10.1103/PhysRevLett.105.246804} {\bibfield  {journal} {\bibinfo  {journal}
  {Phys. Rev. Lett.}\ }\textbf {\bibinfo {volume} {105}},\ \bibinfo {pages}
  {246804} (\bibinfo {year} {2010}{\natexlab{b}})}\BibitemShut {NoStop}%
\bibitem [{\citenamefont {Shi}\ \emph {et~al.}(2013)\citenamefont {Shi},
  \citenamefont {Simmons}, \citenamefont {Ward}, \citenamefont {Prance},
  \citenamefont {Mohr}, \citenamefont {Koh}, \citenamefont {Gamble},
  \citenamefont {Wu}, \citenamefont {Savage}, \citenamefont {Lagally},
  \citenamefont {Friesen}, \citenamefont {Coppersmith},\ and\ \citenamefont
  {Eriksson}}]{ShiPRB2013}%
  \BibitemOpen
  \bibfield  {author} {\bibinfo {author} {\bibfnamefont {Zhan}\ \bibnamefont
  {Shi}}, \bibinfo {author} {\bibfnamefont {C~B}\ \bibnamefont {Simmons}},
  \bibinfo {author} {\bibfnamefont {Daniel~R}\ \bibnamefont {Ward}}, \bibinfo
  {author} {\bibfnamefont {J~R}\ \bibnamefont {Prance}}, \bibinfo {author}
  {\bibfnamefont {R~T}\ \bibnamefont {Mohr}}, \bibinfo {author} {\bibfnamefont
  {Teck~Seng}\ \bibnamefont {Koh}}, \bibinfo {author} {\bibfnamefont
  {John~King}\ \bibnamefont {Gamble}}, \bibinfo {author} {\bibfnamefont {Xian}\
  \bibnamefont {Wu}}, \bibinfo {author} {\bibfnamefont {D~E}\ \bibnamefont
  {Savage}}, \bibinfo {author} {\bibfnamefont {M~G}\ \bibnamefont {Lagally}},
  \bibinfo {author} {\bibfnamefont {Mark}\ \bibnamefont {Friesen}}, \bibinfo
  {author} {\bibfnamefont {S~N}\ \bibnamefont {Coppersmith}}, \ and\ \bibinfo
  {author} {\bibfnamefont {M~A}\ \bibnamefont {Eriksson}},\ }\bibfield  {title}
  {\enquote {\bibinfo {title} {{Coherent quantum oscillations and echo
  measurements of a Si charge qubit}},}\ }\href {\doibase
  10.1103/PhysRevB.88.075416} {\bibfield  {journal} {\bibinfo  {journal} {Phys.
  Rev. B}\ }\textbf {\bibinfo {volume} {88}},\ \bibinfo {pages} {075416}
  (\bibinfo {year} {2013})}\BibitemShut {NoStop}%
\bibitem [{\citenamefont {Thorgrimsson}\ \emph {et~al.}(2017)\citenamefont
  {Thorgrimsson}, \citenamefont {Kim}, \citenamefont {Yang}, \citenamefont
  {Smith}, \citenamefont {Simmons}, \citenamefont {Ward}, \citenamefont
  {Foote}, \citenamefont {Corrigan}, \citenamefont {Savage}, \citenamefont
  {Lagally}, \citenamefont {Friesen}, \citenamefont {Coppersmith},\ and\
  \citenamefont {Eriksson}}]{ThorgrimssonNPJQuantumInfo2017}%
  \BibitemOpen
  \bibfield  {author} {\bibinfo {author} {\bibfnamefont {Brandur}\ \bibnamefont
  {Thorgrimsson}}, \bibinfo {author} {\bibfnamefont {Dohun}\ \bibnamefont
  {Kim}}, \bibinfo {author} {\bibfnamefont {Yuan-Chi}\ \bibnamefont {Yang}},
  \bibinfo {author} {\bibfnamefont {L~W}\ \bibnamefont {Smith}}, \bibinfo
  {author} {\bibfnamefont {C~B}\ \bibnamefont {Simmons}}, \bibinfo {author}
  {\bibfnamefont {Daniel~R}\ \bibnamefont {Ward}}, \bibinfo {author}
  {\bibfnamefont {Ryan~H}\ \bibnamefont {Foote}}, \bibinfo {author}
  {\bibfnamefont {J}~\bibnamefont {Corrigan}}, \bibinfo {author} {\bibfnamefont
  {D~E}\ \bibnamefont {Savage}}, \bibinfo {author} {\bibfnamefont {M~G}\
  \bibnamefont {Lagally}}, \bibinfo {author} {\bibfnamefont {Mark}\
  \bibnamefont {Friesen}}, \bibinfo {author} {\bibfnamefont {S~N}\ \bibnamefont
  {Coppersmith}}, \ and\ \bibinfo {author} {\bibfnamefont {M~A}\ \bibnamefont
  {Eriksson}},\ }\bibfield  {title} {\enquote {\bibinfo {title} {{Extending the
  coherence of a quantum dot hybrid qubit}},}\ }\href {\doibase
  10.1038/s41534-017-0034-2} {\bibfield  {journal} {\bibinfo  {journal} {npj
  Quantum Inf.}\ }\textbf {\bibinfo {volume} {3}},\ \bibinfo {pages} {32}
  (\bibinfo {year} {2017})}\BibitemShut {NoStop}%
\bibitem [{\citenamefont {Széchenyi}\ and\ \citenamefont
  {Pályi}(2019)}]{SzechenyiArxiv2019}%
  \BibitemOpen
  \bibfield  {author} {\bibinfo {author} {\bibfnamefont {Gábor}\ \bibnamefont
  {Széchenyi}}\ and\ \bibinfo {author} {\bibfnamefont {András}\ \bibnamefont
  {Pályi}},\ }\href@noop {} {\enquote {\bibinfo {title} {{Parity-to-charge
  conversion for readout of topological Majorana qubits}},}\ } (\bibinfo {year}
  {2019}),\ \Eprint {http://arxiv.org/abs/1909.02326} {arXiv:1909.02326
  [cond-mat.mes-hall]} \BibitemShut {NoStop}%
\bibitem [{\citenamefont {Walter}\ \emph {et~al.}(2017)\citenamefont {Walter},
  \citenamefont {Kurpiers}, \citenamefont {Gasparinetti}, \citenamefont
  {Magnard}, \citenamefont {Poto{\v{c}}nik}, \citenamefont {Salath{\'{e}}},
  \citenamefont {Pechal}, \citenamefont {Mondal}, \citenamefont {Oppliger},
  \citenamefont {Eichler},\ and\ \citenamefont {Wallraff}}]{WalterPRApp2017}%
  \BibitemOpen
  \bibfield  {author} {\bibinfo {author} {\bibfnamefont {T}~\bibnamefont
  {Walter}}, \bibinfo {author} {\bibfnamefont {P}~\bibnamefont {Kurpiers}},
  \bibinfo {author} {\bibfnamefont {S}~\bibnamefont {Gasparinetti}}, \bibinfo
  {author} {\bibfnamefont {P}~\bibnamefont {Magnard}}, \bibinfo {author}
  {\bibfnamefont {A.}~\bibnamefont {Poto{\v{c}}nik}}, \bibinfo {author}
  {\bibfnamefont {Y}~\bibnamefont {Salath{\'{e}}}}, \bibinfo {author}
  {\bibfnamefont {M}~\bibnamefont {Pechal}}, \bibinfo {author} {\bibfnamefont
  {M}~\bibnamefont {Mondal}}, \bibinfo {author} {\bibfnamefont {M}~\bibnamefont
  {Oppliger}}, \bibinfo {author} {\bibfnamefont {C}~\bibnamefont {Eichler}}, \
  and\ \bibinfo {author} {\bibfnamefont {A}~\bibnamefont {Wallraff}},\
  }\bibfield  {title} {\enquote {\bibinfo {title} {{Rapid High-Fidelity
  Single-Shot Dispersive Readout of Superconducting Qubits}},}\ }\href
  {\doibase 10.1103/PhysRevApplied.7.054020} {\bibfield  {journal} {\bibinfo
  {journal} {Phys. Rev. Appl.}\ }\textbf {\bibinfo {volume} {7}},\ \bibinfo
  {pages} {054020} (\bibinfo {year} {2017})}\BibitemShut {NoStop}%
\bibitem [{\citenamefont {Shevchenko}\ \emph {et~al.}(2009)\citenamefont
  {Shevchenko}, \citenamefont {Ashhab},\ and\ \citenamefont
  {Nori}}]{ShevchenkoReview}%
  \BibitemOpen
  \bibfield  {author} {\bibinfo {author} {\bibfnamefont {Sergey~N.}\
  \bibnamefont {Shevchenko}}, \bibinfo {author} {\bibfnamefont
  {S.}~\bibnamefont {Ashhab}}, \ and\ \bibinfo {author} {\bibfnamefont
  {Franco}\ \bibnamefont {Nori}},\ }\bibfield  {title} {\enquote {\bibinfo
  {title} {{Landau-Zener-Stuckelberg interferometry}},}\ }\href {\doibase
  10.1016/j.physrep.2010.03.002} {\bibfield  {journal} {\bibinfo  {journal}
  {Phys. Rep.}\ }\textbf {\bibinfo {volume} {492}},\ \bibinfo {pages} {1--30}
  (\bibinfo {year} {2009})},\ \Eprint {http://arxiv.org/abs/0911.1917}
  {0911.1917} \BibitemShut {NoStop}%
\bibitem [{\citenamefont {Shirley}(1965)}]{ShirleyPhysRev1965}%
  \BibitemOpen
  \bibfield  {author} {\bibinfo {author} {\bibfnamefont {Jon~H}\ \bibnamefont
  {Shirley}},\ }\bibfield  {title} {\enquote {\bibinfo {title} {{Solution of
  the Schr{\"{o}}dinger Equation with a Hamiltonian Periodic in Time}},}\
  }\href {\doibase 10.1103/PhysRev.138.B979} {\bibfield  {journal} {\bibinfo
  {journal} {Phys. Rev.}\ }\textbf {\bibinfo {volume} {138}},\ \bibinfo {pages}
  {B979--B987} (\bibinfo {year} {1965})}\BibitemShut {NoStop}%
\bibitem [{\citenamefont {Romh{\'{a}}nyi}\ \emph {et~al.}(2015)\citenamefont
  {Romh{\'{a}}nyi}, \citenamefont {Burkard},\ and\ \citenamefont
  {P{\'{a}}lyi}}]{RomhanyiPRB2015}%
  \BibitemOpen
  \bibfield  {author} {\bibinfo {author} {\bibfnamefont {Judit}\ \bibnamefont
  {Romh{\'{a}}nyi}}, \bibinfo {author} {\bibfnamefont {Guido}\ \bibnamefont
  {Burkard}}, \ and\ \bibinfo {author} {\bibfnamefont {Andr{\'{a}}s}\
  \bibnamefont {P{\'{a}}lyi}},\ }\bibfield  {title} {\enquote {\bibinfo {title}
  {{Subharmonic transitions and Bloch-Siegert shift in electrically driven spin
  resonance}},}\ }\href {\doibase 10.1103/PhysRevB.92.054422} {\bibfield
  {journal} {\bibinfo  {journal} {Phys. Rev. B}\ }\textbf {\bibinfo {volume}
  {92}},\ \bibinfo {pages} {054422} (\bibinfo {year} {2015})}\BibitemShut
  {NoStop}%
\bibitem [{\citenamefont {Esterli}\ \emph {et~al.}(2019)\citenamefont
  {Esterli}, \citenamefont {Otxoa},\ and\ \citenamefont
  {Gonzalez-Zalba}}]{EsterliAPL2019}%
  \BibitemOpen
  \bibfield  {author} {\bibinfo {author} {\bibfnamefont {M.}~\bibnamefont
  {Esterli}}, \bibinfo {author} {\bibfnamefont {R.~M.}\ \bibnamefont {Otxoa}},
  \ and\ \bibinfo {author} {\bibfnamefont {M.~F.}\ \bibnamefont
  {Gonzalez-Zalba}},\ }\bibfield  {title} {\enquote {\bibinfo {title}
  {Small-signal equivalent circuit for double quantum dots at
  low-frequencies},}\ }\href {\doibase 10.1063/1.5098889} {\bibfield  {journal}
  {\bibinfo  {journal} {Applied Physics Letters}\ }\textbf {\bibinfo {volume}
  {114}},\ \bibinfo {pages} {253505} (\bibinfo {year} {2019})}\BibitemShut
  {NoStop}%
\bibitem [{\citenamefont {D'Anjou}\ and\ \citenamefont
  {Burkard}(2019)}]{DAnjouPRB2019}%
  \BibitemOpen
  \bibfield  {author} {\bibinfo {author} {\bibfnamefont {B.}~\bibnamefont
  {D'Anjou}}\ and\ \bibinfo {author} {\bibfnamefont {Guido}\ \bibnamefont
  {Burkard}},\ }\bibfield  {title} {\enquote {\bibinfo {title} {Optimal
  dispersive readout of a spin qubit with a microwave resonator},}\ }\href
  {\doibase 10.1103/PhysRevB.100.245427} {\bibfield  {journal} {\bibinfo
  {journal} {Phys. Rev. B}\ }\textbf {\bibinfo {volume} {100}},\ \bibinfo
  {pages} {245427} (\bibinfo {year} {2019})}\BibitemShut {NoStop}%
\end{thebibliography}%
%    \end{thebibliography}
\end{document}